\documentclass[journal]{IEEEtran}

\usepackage[utf8]{inputenc}
\usepackage[T1]{fontenc}
\usepackage[english]{babel}
\usepackage{amsmath}
\usepackage{amssymb}
\usepackage{graphicx}
\usepackage{booktabs}
\usepackage{caption}
\usepackage{subcaption}
\usepackage{xcolor}
\usepackage[hidelinks]{hyperref}
\usepackage{cite}
\usepackage{multirow}
\usepackage{microtype}
\microtypesetup{protrusion=true,expansion=true,tracking=false}
\usepackage{tikz}
\usetikzlibrary{arrows.meta,positioning,shapes.geometric,calc,fit,backgrounds}
\usepackage{algorithm}
\usepackage{algpseudocode}
\usepackage{comment}
\graphicspath{{figs/}}

\usepackage[compact]{titlesec}
\titlespacing*{\subsubsection}{0pt}{4pt plus 1pt minus 1pt}{2pt plus 1pt minus 1pt}
\titlespacing*{\subsection}{0pt}{6pt plus 1pt minus 1pt}{3pt plus 1pt minus 1pt}
\let\oldthebibliography\thebibliography
\renewcommand{\thebibliography}[1]{%
  \oldthebibliography{#1}%
  \setlength{\itemsep}{0pt}%
  \setlength{\parskip}{0pt}%
}

\newcommand{\degr}{\ensuremath{^{\circ}}}
\newcommand{\phw}{\ensuremath{\varphi_{\mathrm{hw}}}}

\title{Passive AoA Estimation of COTS 5G NR Handsets from SRS: A Practical USRP-B210 Implementation}

\author{Ricardo~N.~Barrios-Mu\~{n}oz, Azim~Akhtarshenas, David Gomez-Barquero, and David L\'{o}pez-P\'{e}rez%
\thanks{Ricardo N. Barrios-Mu\~{n}oz, Azim Akhtarshenas, David Gomez-Barquero, and David L\'{o}pez-P\'{e}rez are with the iTEAM Research Institute, Universitat Polit\`ecnica de Val\`encia (UPV), 46022 Valencia, Spain (e-mail: rnbarmuo@teleco.upv.es; aakhtar@doctor.upv.es). 

This research was supported by the Generalitat Valenciana, Spain, through the iTENTE project. 

The authors thank Prof. Narcis Cardona and Juan Gallego Luchoro (Universitat Politècnica de València) for their technical support in fabricating the antenna mast and 3D-printing custom mechanical components during the AoA measurement campaign.}%
}

\begin{document}

\markboth{Journal draft - AoA of COTS 5G handsets from uplink SRS}
{Barrios-Mu\~{n}oz \MakeLowercase{\textit{et al.}}}

\maketitle

\begin{abstract}
This paper presents a passive network-side framework for estimating the uplink angle of arrival (AoA) of unmodified commercial 5G handsets from native Sounding Reference Signal (SRS) transmissions. The proposed system operates on an srsRAN Project gNB and estimates AoA directly from the per-antenna SRS channel estimates using a calibrated two-element Universal Software Radio Peripheral (USRP)~B210 receiver and a MUSIC-based estimator, requiring neither protocol modifications nor user-equipment cooperation. Independent SRS channel estimates enable simultaneous AoA estimation for up to four commercial handsets.
The framework is evaluated through indoor and outdoor measurement campaigns. Indoor measurements on bands n40 and n78 achieve a root-mean-square error of $1.5^\circ$ on n40 over a $\pm30^\circ$ broadside sector. Outdoor experiments in a live 5G standalone deployment demonstrate simultaneous multi-user operation and show that estimation accuracy is primarily governed by propagation conditions and received signal-to-interference-plus-noise ratio (SINR) rather than transmitter distance. The simulation results demonstrate the feasibility of passive uplink AoA estimation using native 5G signaling, low-cost SDR hardware, and commercial handsets.
\end{abstract}



\section{Introduction}

Angle of arrival (AoA) provides spatial information about wireless transmitters by exploiting the phase differences measured across a coherent antenna array,
with applications ranging from localization and tracking to trajectory planning and flight optimization for aerial base stations (BSs)~\cite{akhtarshenas2026centralized, akhtarshenas2026wind}.
In 5G NR, the uplink Sounding Reference Signal (SRS) provides a natural basis for AoA estimation because the gNB already computes per-antenna channel estimates for communication purposes.
Recent work has demonstrated SRS-based AoA estimation using SDR platforms and calibrated antenna arrays~\cite{spanos2024angle,hu2023srs,wachowiak2022angle,ceresoli2025aoa},
reflecting the broader trend toward integrated sensing and communication (ISAC), 
where communication signals are reused for sensing without dedicated waveforms~\cite{gonzalez2024integrated}.
Despite this progress, most existing implementations rely on dedicated transmitters, modified software stacks, custom processing pipelines, or controlled laboratory environments
Passive AoA estimation from unmodified commercial handsets using the native SRS processing of an open-source 5G gNB remains largely unexplored, particularly for simultaneous estimation of multiple user equipment (UE) under realistic propagation conditions.

In this work, we address this gap by developing a network-side AoA framework based on an unmodified srsRAN Project gNB and a coherent two-element Universal Software Radio Peripheral (USRP)~B210 receiver.

\subsection{Related Work}\label{sec:lit}

We place this work against four strands: 
uplink SRS/CSI AoA estimators, 
SDR and USRP AoA testbeds, 
open-source 5G positioning stacks,
and passive and COTS-UE cellular sensing. 

\noindent\textit{1) Uplink SRS and CSI AoA estimators:} 
They exploit uplink reference signals to estimate the direction of arrival using array signal processing techniques.
Spanos \emph{et al.}~\cite{spanos2024angle} estimate UE AoAs from uplink SRS, 
comparing MUSIC, ESPRIT and JADE-ESPRIT algorithms on a dedicated three-element ULA. 
Hu \emph{et al.}~\cite{hu2023srs} take SRSs from commercial devices and calibrate the channel frequency response,
reaching $\approx5\degr$ single-path and $\approx10\degr$ indoor-multipath error (95th pct.) - the latter matches the multipath spread we report in this paper. 
Liu \emph{et al.}~\cite{liu2025model} correct the angle-dependent phase error of a gNB uplink array with a model-driven neural network, 
and Shah \emph{et al.}~\cite{shah2025evaluation} contrast uplink SRS-based and downlink positioning reference signals (PRSs)-based positioning under Line-of-Sight (LoS) and Non-Line-of-Sight (NLoS),
quantifying the NLOS penalty. 
All these studies use a purpose-built SRS transmitter, a large array, or a network-side export; 
none reads the AoA of unmodified handsets straight from a stock gNB.

\noindent\textit{2) SDR and USRP AoA testbeds:}
They experimentally validate AoA algorithms while addressing antenna calibration, synchronization, and hardware impairments.
Wachowiak and Kryszkiewicz~\cite{wachowiak2022angle} developed an SDR-based AoA estimation system using a USRP B210 and the Root-MUSIC algorithm. 
They analyzed hardware synchronization errors and multipath effects, 
proposed calibration and signal-processing techniques to mitigate them, 
and validated the system experimentally with a publicly available implementation.
Sheremet and Fokin~\cite{sheremet2026experimental,sheremet2024campus} combine two B210s into a four-element uniform linear array (ULA), 
and separately report that a minimalist two-element array is sensitive to reflections at wide angles, 
corroborating some findings in our experimental results.
Xhafa \emph{et al.}~\cite{xhafa2025experimental} give a full experimental characterisation of how RX-chain and antenna-array calibration bound AoA accuracy on a USRP/N310 array.
Ceresoli \emph{et al.}~\cite{ceresoli2025aoa} frame uplink-SRS AoA as a network-native RAN service with a repeatable calibration methodology.
These testbeds establish effective calibration procedures,
but they rely on larger antenna arrays and dedicated or emulated transmitters, 
rather than passive two-element reception of live commercial off-the-shelf (COTS) signals.

\noindent\textit{3) Open-source 5G positioning stacks:} 
They implement AoA estimation using native NR signaling on platforms such as OpenAirInterface (OAI) and srsRAN.
Palam\`a \emph{et al.}~\cite{palama2024positioning} survey SDR-based 5G positioning solutions,
The OAI ecosystem supports uplink SRS-based Time Difference of Arrival (TDoA) positioning~\cite{ahadi2023srs}, 
as well as a complete open-source UL-TDoA framework with NRPPa/LMF signalling~\cite{malik2024concept}.
The srsRAN ecosystem supports similar features. 
However, unlike OAI, srsRAN does not natively support PRSs or standardized positioning measurements.
Thus, in this paper, 
rather than relying on standardized positioning reports, 
our anchor extracts the per-antenna SRS channel estimates directly from the srsRAN PHY layer.

\noindent\textit{4) Passive and COTS-UE cellular sensing:}
They estimates the AoA of commercial cellular transmissions without modifying the network or requiring user-equipment cooperation.
Huang \emph{et al.}~\cite{huang2026fuse} localize UAVs passively from multi-UE SRS echoes, consdering  frequency and amplitude impairments and correcting per-UE timing.
Gangula \emph{et al.}~\cite{gangula2025bistatic} show bistatic sensing on the standard uplink without PHY changes, 
and Jopanya and Osorio~\cite{jopanya2025utilizing} detect drones passively from downlink SSB grids. 
These confirm an unmodified handset's uplink carries exploitable spatial information - the premise our anchor rests on - but pursue detection or CSI fingerprints rather than a per-UE AoA.

\begin{table*}[!t]
    \centering
    \caption{Experimental uplink AoA systems closest to this work. Passive
    means no dedicated transmitter, no RAN modification and no UE
    cooperation. n/r = not reported.}
    \label{tab:prior_art}
    \footnotesize
        \begin{tabular}{@{}p{3.2cm}p{2.9cm}p{2.9cm}p{2.5cm}cp{3.4cm}@{}}
        \toprule
        Work & Array & Source & Environment & Passive? & Reported accuracy \\
        \midrule
        Hu \emph{et al.}~\cite{hu2023srs} & gNB array (n/r) & COTS handset SRS &
        Indoor & No & $\approx5\degr$ single-path, $\approx10\degr$ multipath
        (95th pct.) \\
        Wachowiak and Kryszkiewicz~\cite{wachowiak2022angle} & B210, 2-element
        ULA & Dedicated Tx, wired calibration tone & n/r & No & n/r \\
        Sheremet and
        Fokin~\cite{sheremet2026experimental,sheremet2024campus} & 2- to
        4-element ULA (B210 pair, ext.\ sync) & n/r & Campus & No & MUSIC
        std.\ dev.\ $<0.4\degr$ (4-element) \\
        Ceresoli \emph{et al.}~\cite{ceresoli2025aoa} & N310, 3-element ULA &
        Dedicated USRP Tx & n/r & No & n/r \\
        Ge \emph{et al.}~\cite{ge2022experimental,ge2023experimental} & mmWave
        single BS & n/r & n/r & No & $\approx4.6\degr$ \\
        \midrule
        This work & B210, 2-element ULA & COTS handsets (up to 4 in parallel) &
        Indoor + outdoor, live network & Yes & $1.5\degr$ RMS indoor broadside
        sweep (indicative); $3$-$10\degr$ outdoor \\
        \bottomrule
        \end{tabular}
\end{table*}

\subsection{Novelty and Contributions}

Table~\ref{tab:prior_art} compares the closest experimental uplink AoA implementations. 
Existing work relies on calibrated SDR arrays or modified network infrastructure.
Motivated by~\cite{wachowiak2022angle},
our study performs passive AoA directly from the native SRS channel estimates of an unmodified srsRAN gNB using commercial handsets.
Therefore, the main contributions are as follows:

\begin{itemize}
    \item \textbf{Passive network-side AoA estimation.} 
    We exploit native per-antenna SRS channel estimates from the srsRAN PHY without protocol modifications, dedicated positioning functions, or UE cooperation.
    
    \item \textbf{Multi-UE AoA estimation.} 
    A two-element USRP~B210 estimates the AoA of up to four simultaneous COTS handsets by exploiting UE-specific SRS resources and RNTI-based identification.
    
    \item \textbf{Experimental validation.} 
    We validate the approach in indoor and outdoor 5G~standalone (SA) deployments, 
    demonstrating that passive SRS-based AoA is feasible with commercial handsets and that its accuracy is primarily governed by propagation conditions and received SINR rather than distance.
\end{itemize}


\section{SYSTEM DESIGN AND METHODOLOGY}\label{sec:sys}

\begin{algorithm}[t]
\caption{MUSIC-Based AoA Estimation}
\label{alg:music}
\begin{algorithmic}[1]
\Require SRS channel estimates $c_0,c_1$
\Ensure Estimated AoA $\hat{\theta}_{\mathrm{axis}}$

\State Compute the averaged inter-element phase:
\Statex \hspace{\algorithmicindent}
$\Delta\varphi \gets
\angle\!\left(\left\langle c_0c_1^*\right\rangle\right)$

\State Obtain a coarse AoA from the phase-interferometer relation:
\Statex \hspace{\algorithmicindent}
$\theta_{\mathrm{coarse}} \gets
\arcsin\!\left(-\frac{\lambda\Delta\varphi}{2\pi d}\right)$

\State Form the spatial covariance matrix:
\Statex \hspace{\algorithmicindent}
$\mathbf{R} \gets \frac{1}{N}\mathbf{X}\mathbf{X}^{H}$

\State Extract the noise subspace $\mathbf{E}_n$ from $\mathbf{R}$

\State Evaluate the MUSIC pseudospectrum:
\Statex \hspace{\algorithmicindent}
$P(\theta) \gets
\left|\mathbf{a}^{H}(\theta)\mathbf{E}_n\right|^{-2}$

\State Refine the AoA within a $\pm15^\circ$ search window:
\Statex \hspace{\algorithmicindent}
$\hat{\theta} \gets
\displaystyle\arg\max_{\theta\in
[\theta_{\mathrm{coarse}}-15^\circ,\,
\theta_{\mathrm{coarse}}+15^\circ]} P(\theta)$

\State Convert the estimated angle to the array-axis convention:
\Statex \hspace{\algorithmicindent}
$\hat{\theta}_{\mathrm{axis}} \gets 90^\circ-\hat{\theta}$

\State \Return $\hat{\theta}_{\mathrm{axis}}$
\end{algorithmic}
\end{algorithm}

\subsection{System Setup}

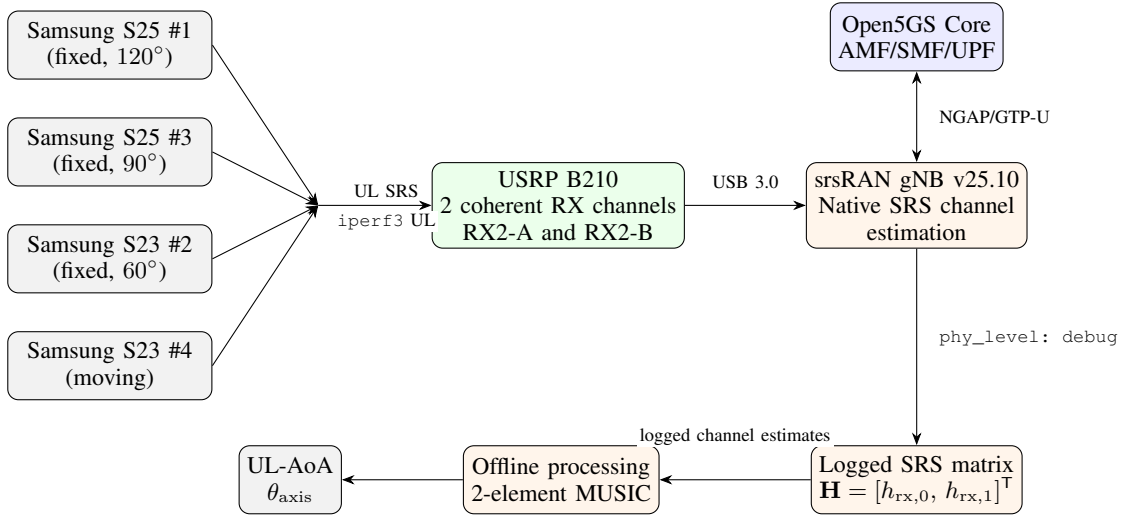
\begin{figure*}[!t]
\centering
\makebox[\textwidth][c]{%
\begin{tikzpicture}[
  node distance=9mm and 14mm,
  box/.style={
    draw,
    rounded corners,
    align=center,
    minimum height=9mm,
    inner sep=3pt,
    font=\small
  },
  ue/.style={box,fill=gray!10,minimum width=27mm},
  rf/.style={box,fill=green!8},
  sw/.style={box,fill=orange!8},
  core/.style={box,fill=blue!8},
  >={Stealth}
]

\node[ue] (ue1) {Samsung S25 \#1\\(fixed, $120^\circ$)};
\node[ue,below=5mm of ue1] (ue2) {Samsung S25 \#3\\(fixed, $90^\circ$)};
\node[ue,below=5mm of ue2] (ue3) {Samsung S23 \#2\\(fixed, $60^\circ$)};
\node[ue,below=5mm of ue3] (ue4) {Samsung S23 \#4\\(moving)};

\coordinate (comb) at ($(ue1.east)!0.5!(ue4.east) + (14mm,0)$);

\node[rf,right=15mm of comb] (b210)
  {USRP B210\\2 coherent RX channels\\RX2-A and RX2-B};

\node[sw,right=16mm of b210] (phy)
  {srsRAN gNB v25.10\\Native SRS channel\\estimation};

\node[core,above=12mm of phy] (core)
  {Open5GS Core\\AMF/SMF/UPF};

\node[sw,below=26mm of phy] (log)
  {Logged SRS matrix\\$\mathbf{H}=[h_{\mathrm{rx},0},\,h_{\mathrm{rx},1}]^{\mathsf T}$};

\node[sw,left=20mm of log] (aoa)
  {Offline processing\\2-element MUSIC};

\node[box,fill=gray!10,left=16mm of aoa] (out)
  {UL-AoA\\$\theta_{\mathrm{axis}}$};

\draw[->] (ue1.east) -- (comb);
\draw[->] (ue2.east) -- (comb);
\draw[->] (ue3.east) -- (comb);
\draw[->] (ue4.east) -- (comb);

\draw[->] (comb) -- (b210.west)
  node[pos=.6,above=2pt,font=\scriptsize,fill=white,inner sep=1pt]{UL SRS}
  node[pos=.6,below=2pt,font=\scriptsize,fill=white,inner sep=1pt]{\texttt{iperf3} UL};

\draw[->] (b210) -- (phy)
  node[midway,above=3pt,font=\scriptsize]{USB 3.0};

\draw[<->] (phy) -- (core)
  node[midway,right=5pt,font=\scriptsize]{NGAP/GTP-U};

\draw[->] (phy) -- (log)
  node[pos=.45,right=5pt,font=\scriptsize]{\texttt{phy\_level: debug}};

\draw[->] (log) -- (aoa)
  node[midway,above=12pt,font=\scriptsize,fill=white,inner sep=1pt]
  {logged channel estimates};

\draw[->] (aoa) -- (out);

\end{tikzpicture}%
}

\caption{Overall data path for UL-AoA estimation.}
\label{fig:datapath}

\end{figure*}

\begin{figure}[t]
    \centering
    \includegraphics[width=0.85\columnwidth]{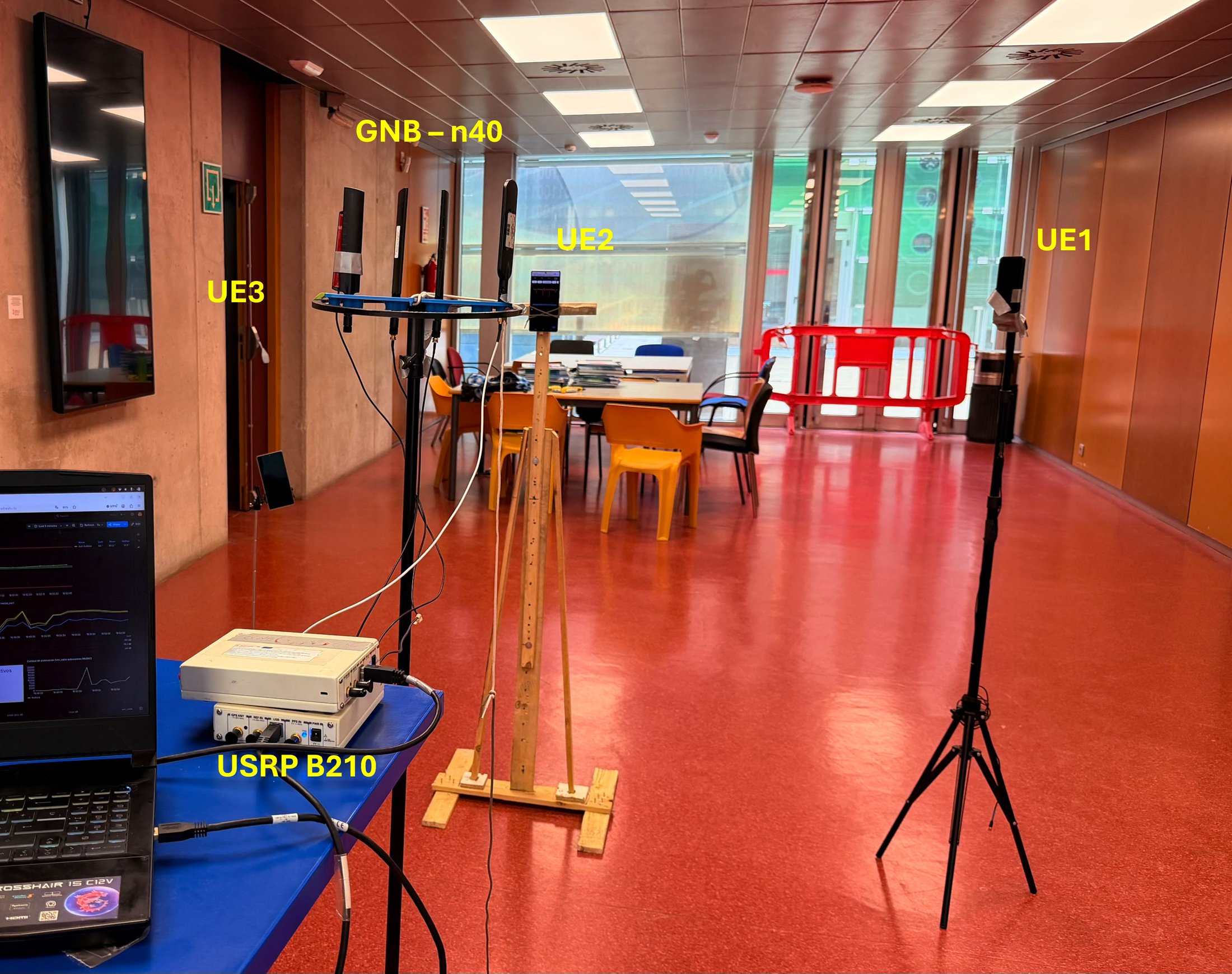}
    \caption{Indoor experimental setup used in Campaign~1.}
    \label{fig:setup2}
\end{figure}

\begin{figure}[t]
\centering

\begin{subfigure}[t]{0.9\columnwidth}
    \centering
    \includegraphics[width=.9\linewidth]{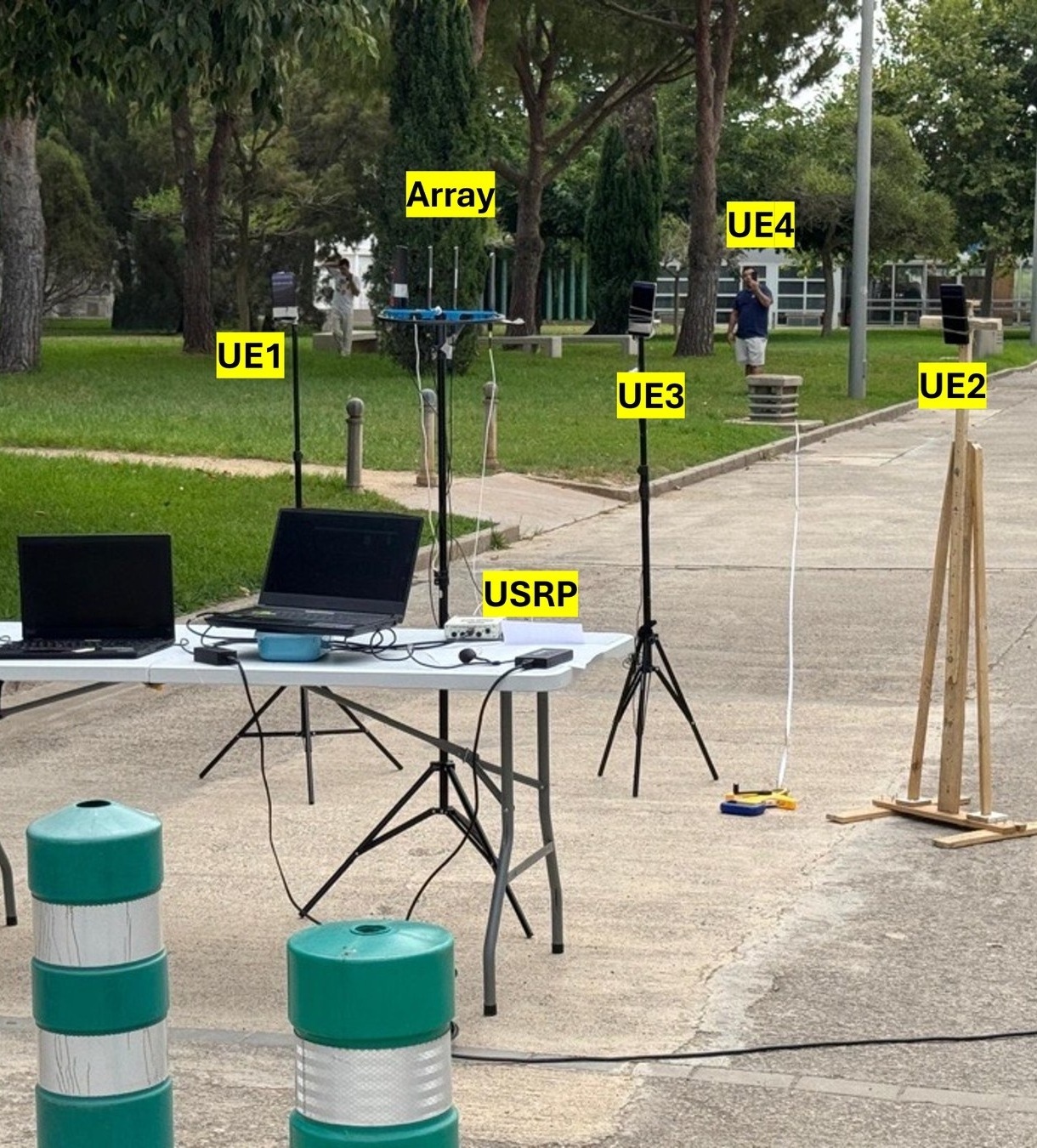}
    \caption{Parallel geometry: array on a tripod, three fixed UE and one real walking person.}
    \label{fig:outdoor_setup_a}
\end{subfigure}

\vspace{0.5em}

\begin{subfigure}[t]{0.8\columnwidth}
    \centering
    \includegraphics[width=1\linewidth]{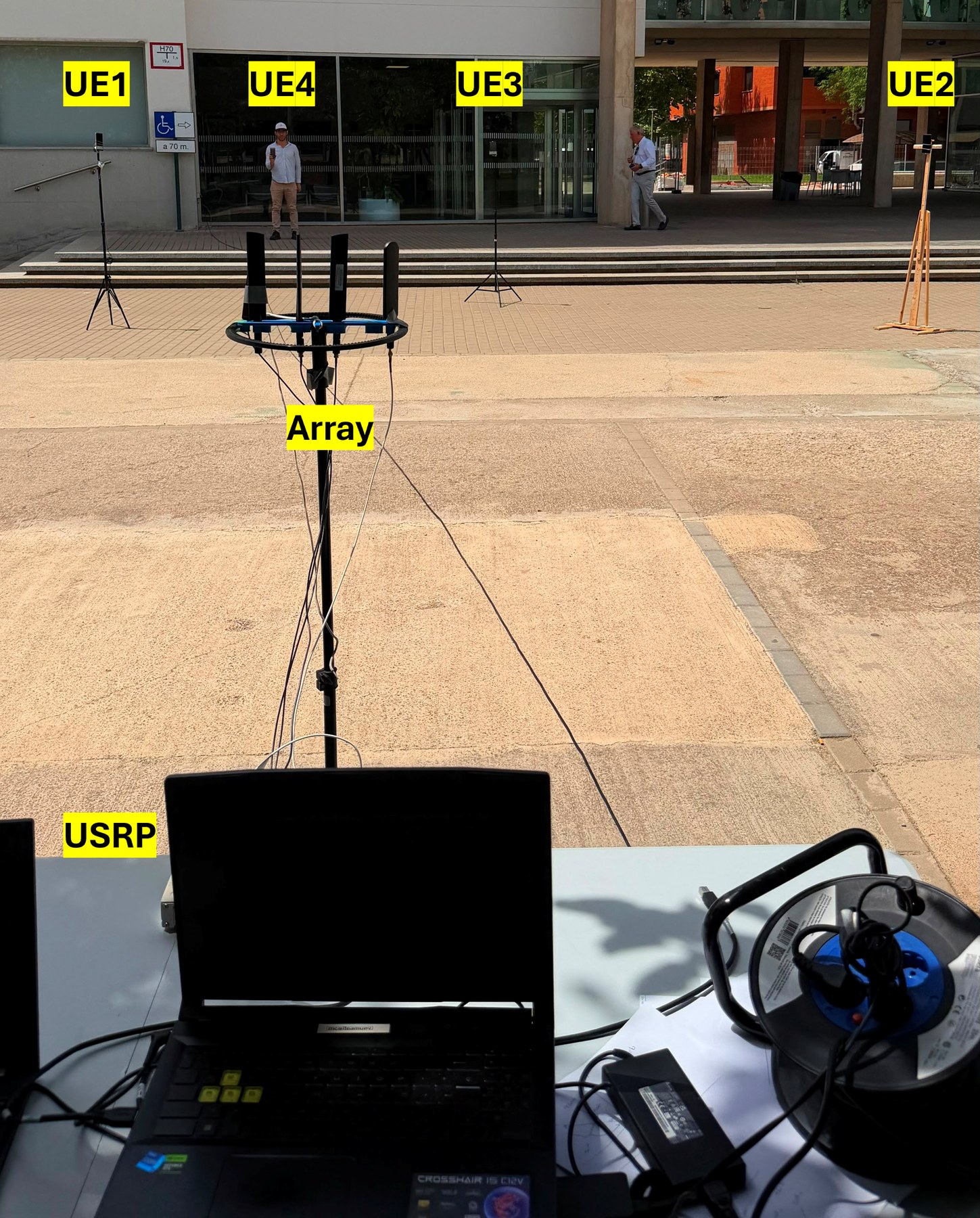}
    \caption{Face-to-Face geometry: Three fixed UEs and one real walking person 0º to 180º.}
    \label{fig:outdoor_setup_b}
\end{subfigure}

\caption{Outdoor measurement setups used in Campaign~2.}
\label{fig:outdoor_setup}

\end{figure}

The proposed experimental setup consists of commercial COTS 5G UE, a single USRP~B210 software-defined radio, an srsRAN Project gNB (version 25.10), and an Open5GS 5G SA core network (See Fig.~\ref{fig:datapath}).
The commercial handsets generate continuous uplink traffic using \texttt{iperf3}, 
while periodic SRS resources are gathered at the gNB.
The USRP~B210 is controlled directly by the srsRAN gNB via USB~3.0, 
which estimates and records the native per-antenna SRS channel matrix for each sounding occasion. 
The recorded channel estimates are then processed off-line using a two-element MUSIC algorithm to recover the UL-AoA. 
Since AoA estimation operates solely on the logged channel estimates,
no additional radio access, parallel SDR receiver, or duplicate USB interface is required.
The proposed system was evaluated in both indoor (see Fig.~\ref{fig:setup2}) and outdoor environments (see Fig.~\ref{fig:outdoor_setup}-a and Fig.~\ref{fig:outdoor_setup}-b) to assess its performance under diverse propagation conditions. 
The indoor experiments were conducted in the vicinity of a small cell, 
whereas the outdoor experiments considered two challenging propagation scenarios: 
(i) measurements near a glass wall, 
where strong multipath propagation can arise due to reflections, 
and (ii) measurements at the cell edge, 
where ISI may be more pronounced due to the propagation environment. 
The common hardware and software configurations are described in this section, 
while the specific experimental setups and measurement procedures for the indoor and outdoor scenarios are presented in Sections~\ref{indoor_indoor} and~\ref{outdoor_outdoor}, respectively.

\noindent\textit{1) UE:}\label{ue_ue}
We used commercial Samsung 5G handsets.
The indoor campaign used three static UEs (UE1-UE3), 
while the outdoor campaign added UE4 as a moving reference, 
walked across $0$-$180\degr$ and displaced along the $2$-$61$\,m range 
(see Table~\ref{tab:geometry}). 
Since RNTIs may change when a UE re-attaches different cells, 
we associate each SRS occasion with the UE identified by the RNTI active at that time.
A 3D-printed circular goniometer,
mounted at the array and aligned to boresight, 
sets each static UE's true AoA with an error about $\pm1$-$2\degr$.

\noindent\textit{2) SDR:}\label{2T2R}
The proposed platform employs a single USRP~B210 (serial \texttt{30F7DC2})~\cite{b210} based on one Analog Devices AD9361~\cite{ad9361} transceiver and configured in a 2T2R mode. 
The downlink remains a conventional rank-1 transmission; 
the second receive chain is therefore solely used in our setup to enable AoA estimation using a coherent two-element antenna array.
Using a single B210 rather than multiple ones,
receive chains share the same local oscillator (LO), 
giving a stable inter-channel phase relationship up to a fixed hardware offset $\phi_{\mathrm{hw}}$,
which can be mitigated using a one-time calibration (Section~\ref{sec:calib})  and avoids the continuous phase drift of independent LOs in separate radios.
\noindent\textit{3) Antenna array:}\label{antenna}
The array connected to the SDR uses two \emph{identical} wideband, vertically polarised and omnidirectional Poynting OMNI-A0085 antennas~\cite{poynting},
covering 617-3800\,MHz with a peak gain of 3.5\,dBi over 1710-2700\,MHz (n40) and 2.5\,dBi over 3400-3800\,MHz (n78), 
so one antenna type covers both campaign bands with only 1\,dB of gain difference.
We space the two antennas $d=\lambda/2$ ($6.39$\,cm at n40 and $4.16$\,cm at n78). 
Because \texttt{tx\_mode=continuous} makes srsRAN receive on the B210 RX2 ports, 
we wire the array to RX2-A (ch0) and RX2-B (ch1), 
exactly where we measured $\phw$ in \eqref{nonono},
so the calibration transfers directly. 
The array sits on a tripod at 1.5\,m height.\\
\begin{table*}[tp]
    \centering
    \caption{Experimental geometry for the indoor (Campaign~1) and outdoor (Campaign~2) measurements.}
    \label{tab:geometry}
    \footnotesize
        \begin{tabular}{@{}cllcccc@{}}
        \toprule
        Campaign & Node & Model & Height (m) & Angle & Distance (m) & Role \\
        \midrule
        \multirow{4}{*}{Indoor}
        & Array (gNB) & USRP B210 (RX2-A/B)       & 1.5     & --               & --         & Receiver \\
        & UE1         & Samsung S23 (SM-S911B)    & 1.5     & Elev.\ $0^\circ$ & 2          & Static \\
        & UE2         & Samsung S25 Ultra (SM-S938B) & 1.5  & Elev.\ $0^\circ$ & 2/4/6      & Movable \\
        & UE3         & Samsung S25 Ultra (SM-S938B) & 1.0  & Elev.\ $14^\circ$& 2          & Static \\
        \midrule
        \multirow{5}{*}{Outdoor}
        & Array (gNB) & USRP B210 (RX2-A/B)       & 1.5/2   & --               & --         & Receiver \\
        & UE1         & Samsung S23 (SM-S911B)    & 1.5/2   & Az.\ $120^\circ$ & 11.5       & Static \\
        & UE2         & Samsung S25 Ultra (SM-S938B) & 1.5/2 & Az.\ $60^\circ$  & 11.5       & Static \\
        & UE3         & Samsung S25 Ultra (SM-S938B) & 1.5/2 & Az.\ $90^\circ$  & 11.5       & Static (boresight) \\
        & UE4         & Samsung S23 (SM-S911B)    & 2.0     & Moving           & 2/4/27/42/61 & Walking \\
        \bottomrule
        \end{tabular}
\end{table*}



\noindent\textit{4) UE-side monitoring and uplink traffic:}\label{sec:traffic}
Each handset runs an in-house Android drive-test app, 
logging at about one sample per second the serving-cell RSRP, RSRQ, SINR, DL/UL throughput, ARFCN, band, PCI and GPS position, 
alongside scripted per-block results (iperf uplink Mbps, ping, jitter, video). 
This is our UE-side ground truth, 
joined to the gNB-side SRS data by absolute timestamp. 
We generate the uplink traffic with iperf3 servers on the core gateway (10.45.0.1): 
one port per handset (5201 single-UE, 5201-5204 multi-UE), 
each running looped 15\,s uplink blocks so that every SRS occasion used for AoA coincides with representative uplink activity.
\noindent\textit{5) 5G Core Network (Open5GS):}\label{sec:core}
The gNB connects to a 5G core network to provide standard-compliant standalone (SA) operation. 
We employ Open5GS, 
an open-source 5G SA core, 
running on the same host as the srsRAN Project gNB, 
and started first when the gNB establishes an N2 (NGAP) connection with the Access and Mobility Management Function (AMF) during initialization and performs the NG Setup procedure. 
The srsRAN Project implements a 3GPP Release~15/16-compliant NR gNB, 
while Open5GS provides a Release~16 5G core.
The two interoperate through the standard NG interface using the 5G service-based architecture.
Control-plane functions communicate through the Service-Based Interface (SBI),
the Network Repository Function (NRF) provides service registration and discovery, 
and the Service Communication Proxy (SCP) routes SBI traffic among them.
We configured one Public Land Mobile Network (PLMN) 99970 with Mobile Country Code (MCC)~999 and Mobile Network Code (MNC)~70, one tracking area (TAC~1) and one slice (S-NSSAI, SST~1), 
matched by the gNB,
so that NG Setup and UE registration succeed. 
The control plane uses N2 (NGAP) to the AMF at \texttt{127.0.0.5}, the user plane uses GTP-U to the UPF and both run over loopback, 
since the core and gNB share one host. 
The UPF terminates GTP-U and exposes the UE subnet (\texttt{10.45.0.0/16}, gateway \texttt{10.45.0.1}) on the \texttt{ogstun} TUN interface, 
where we ran the iperf3 server that drives the AoA measurement by keeping every UE transmitting.
\noindent\textit{6) gNB Configuration (srsRAN):}\label{sec:gnb}
\begin{table}[tp]
\centering
\setlength{\tabcolsep}{3pt}
\caption{srsRAN gNB configuration for the n40 TDD cell.}
\label{tab:yaml}
\footnotesize
\begin{tabular}{@{}lll@{}}
\toprule
Group & Parameter & Value \\
\midrule
Cell & band & n40 (TDD) \\
Cell & channel\_bandwidth\_MHz & 20\,MHz \\
Cell & common\_scs & 20\,kHz \\
Cell & PCI / PLMN / TAC & 1 / 99970 / 1 \\
Cell & TDD pattern & DDDDDDXUUU (10 slots/5\,ms) \\
Antennas & nof\_antennas\_ul & 2 (B210 constraint) \\
Antennas & nof\_antennas\_dl & 2 (B210 constraint) \\
Antennas & pdsch\_max\_rank & 1 \\
SRS & srs\_period\_ms / srs\_tx\_comb & 10 / 4 \\
SRS & nof\_sym\_per\_resource & 1 \\
SRS & max\_nof\_sym\_per\_slot & 1 \\
RU/SDR & driver / serial & UHD b200 / 30F7DC2 \\
RU/SDR & srate / otw\_format & 23.04\,MHz / sc12 \\
RU/SDR & tx\_gain / rx\_gain & 89 / 28\,dB \\
RU/SDR & clock / tx\_mode & internal / continuous \\
Logging & phy\_level & debug \\
PCAP & mac\_enable / mac\_type & true / DLT \\
\bottomrule
\end{tabular}
\end{table}
A small set of parameters is particularly relevant to the AoA implementation:
\begin{itemize}
    \item \texttt{nof\_antennas\_dl = nof\_antennas\_ul = 2}: 
    AoA estimation requires two uplink receive antennas. 
    The UHD B2x0 validator requires the same number of antennas in the downlink and uplink, 
    so both parameters must be set to 2.
    However, the downlink is restricted to rank~1 using \texttt{pdsch max\_rank=1}. 
    Thus, the 2T2R configuration provides the two receive channels required for AoA estimation without operating as a $2\times2$ MIMO link.

    \item \texttt{mac\_type = dlt}: 
    This setting is required by srsRAN when \texttt{nof\_antennas}~$\geq 2$ and therefore enables the two-antenna configuration required for AoA estimation.

    \item \texttt{otw\_format = sc12}: 
    Two receive channels sampled at 23.04\,MHz exceed the available USB3 bandwidth when using the standard \texttt{sc16} format.
    We therefore use the 12-bit \texttt{sc12} over-the-wire format, 
    which reduces the required data rate while retaining both receive channels.

    \item \texttt{tx\_mode = continuous}: 
    This configuration places the B210 receive chains on the RX2 ports. 
    These are the same ports for which the hardware phase offset $\phw$ was characterised, 
    allowing the previously obtained calibration to be applied directly.

    \item \texttt{srs\_tx\_comb = 4}: 
    This setting provides four orthogonal SRS comb offsets, 
    allowing up to four UEs to transmit SRS simultaneously without mutual interference.

    \item \texttt{phy\_level = debug}: 
    At this log level, the PHY reports the complex SRS channel estimate for each receive antenna at every SRS occasion.
    These per-antenna channel estimates are the inputs used for AoA estimation and are obtained directly from srsRAN without modifying the PHY implementation.
\end{itemize}
Let $\rho$ denote the SNR of the SRS occasion, 
computed over the two antennas:
\begin{equation}
    \rho^2 = \frac{|h_0|^2 + |h_1|^2}{\sigma^2} = 2 \times \mathrm{SNR}_{\mathrm{SRS}}
    \label{eq:rho_snr}
\end{equation}
where $h_0, h_1$ are the raw wideband channel coefficients per RX port,
and $\sigma^2$ is the noise variance that the srsRAN estimates on the same occasion. 
For each SRS occasion,
the debug log records $\mathbf{H} = \rho\,[h_{\mathrm{rx0}};\,h_{\mathrm{rx1}}]$, 
tagged with a timestamp and RNTI,
which are used to attribute occasions to UEs and join them to the UE-side ground truth. 
Since $\rho$ is real-valued, 
it does not affect the inter-antenna phase in Eq.~\eqref{eq:rho_snr}.\\
\noindent\textit{7) Timing, sounding, and measurand definitions:}\label{sec:timing}
The AoA data lives in the uplink SRS, 
so the frame timing decides how often we can sound each UE.
The cell runs TDD with srsRAN's default n40 pattern, \texttt{DDDDDDXUUU}: 
one 5\,ms period holding 10 slots (0.5\,ms each at 30\,kHz SCS), 
split into six downlink slots, one special slot (DL$\to$UL switch and guard period) and three uplink slots, 
which fixes the sounding budget since SRS travels only in the uplink slots. 
We set \texttt{srs\_period\_ms}~$=10$, 
so each UE sounds once every 10\,ms ($\sim$100 occasions/s), 
and each AoA estimate averages over thousands of these occasions, 
suppressing per-occasion noise. 
Comb~4 gives four orthogonal frequency-domain offsets, 
enabling the simultaneous multi-UE measurements.
We run the SDR on its \texttt{internal} clock with no GPSDO: 
a single gNB and one LO feed the two coherent RX ports, 
so no inter-node time synchronization is needed and the shared LO already guarantees phase-coherent capture across the two ports within each occasion.
We align the gNB host clock (UTC) and the UE app clock (local time) for the timestamp join; 
the joined results (SINR gate, UE4 time series, throughput figures) are robust to it. 
The PHY also logs a per-occasion timing alignment estimate \texttt{t\_align} (SRS-derived timing advance).
We define the logged measurands precisely: for every SRS occasion the srsRAN PHY estimates, over the SRS resource
elements, a received power \texttt{rsrp\_db} and a noise variance
\texttt{noise\_var\_db}, both in dB, giving the SRS SINR as their
log-domain difference,
\begin{equation}
\mathrm{SINR}_{\mathrm{SRS}} = \texttt{rsrp\_db} - \texttt{noise\_var\_db},
\label{eq:srssinr}
\end{equation}
exactly as srsRAN logs it per occasion. All confidence intervals reported in this article use the distinct set-point as the unit of replication, where a set-point denotes one UE at one fixed AoA in one scene, accounting for the temporal correlation among SRS occasions within a set-point and quantifying uncertainty across measured set-points rather than across individual occasions.
For each configuration we take
the per-set-point mean errors, average them over the $N$ distinct
set-points, and report a two-sided Student-$t$ $95\%$ confidence
interval with $N-1$ degrees of freedom.
\noindent\textit{8) Angle convention and front/back ambiguity:}
Angles are reported as $\theta_{\mathrm{axis}}\in[0^\circ,180^\circ]$, measured with respect to the array axis, where boresight corresponds to $90^\circ$. Because a two-element ULA provides only one spatial degree of freedom, signals arriving from symmetric directions in front of and behind the array produce identical inter-element phase differences, resulting in the inherent front/back ambiguity; in both measurement campaigns, this is eliminated by restricting the UE locations to the forward half-plane of the array.
\noindent\textit{9) Theoretical error baseline:}\label{sec:crlb}
To establish a noise-limited performance bound, we derive the phase and AoA estimation variance for the two-element interferometer under additive white Gaussian noise. For the $i$th SRS occasion, the received channel estimate on antenna $m\in\{0,1\}$ is modeled as
\begin{equation}
c_{m,i}=a_m s_i+n_{m,i},
\end{equation}
where $a_m$ is the complex channel coefficient, $s_i$ is the transmitted SRS symbol, and $n_{m,i}$ is zero-mean complex Gaussian noise. Assuming identical per-occasion linear SINR $\gamma$ on both receive chains, the averaged cross-correlation over a window of $K$ SRS occasions~\cite{spanos2024angle,hu2023srs},
\begin{equation}
\hat z=\frac{1}{K}\sum_{i=1}^{K} c_{0,i}c_{1,i}^{*},
\end{equation}
has mean $P_s e^{j\Delta\varphi}$ and complex variance
$(2P_sP_n+P_n^2)/K$, where $\varphi$ is the phase of a received signal. Under the small-error approximation, the phase variance becomes 
\begin{equation}
\operatorname{var}(\Delta\varphi)=
\frac{1}{K}\frac{1+\gamma}{\gamma^{2}},
\label{eq:vardphi}
\end{equation}
which approaches the familiar high-SNR limit $1/(K\gamma)$. The exact small-error expression,
$(1/\gamma+1/2\gamma^2)/K$, differs from
Eq.~\eqref{eq:vardphi} by less than 0.35\,dB for SINR above approximately 7\,dB. We therefore use Eq.~\eqref{eq:vardphi} throughout as a conservative upper bound.
The phase variance is converted into AoA variance using the Jacobian of Eq.~\eqref{eq:geom},
\begin{equation}
\sigma_\theta=
\frac{\lambda}
{2\pi d\,|\sin\theta_{\mathrm{axis}}|}
\sqrt{\operatorname{var}(\Delta\varphi)},
\label{eq:sigtheta}
\end{equation}
showing that angular uncertainty is minimized at boresight and increases toward the end-fire directions because of the geometric sensitivity of the interferometer. A numerical Jacobian propagated through the complete estimator agrees with Eq.~\eqref{eq:sigtheta} within 2\%.

\subsection{Calibration}
\label{sec:calib}

Accurate AoA estimation requires compensation of the constant phase mismatch between the two receive chains.

\noindent\textit{1) Hardware phase offset:}
The common LO ensures frequency coherence between the receive channels but does not guarantee zero relative phase: differences in cable lengths and analog front-end paths introduce the hardware phase offset, $\phi_{\mathrm{hw}}$. Before AoA estimation, the second channel is corrected as
\begin{equation}
c_1 \leftarrow c_1 e^{-j\phi_{\mathrm{hw}}},
\qquad
\phi_{\mathrm{hw}}
=
\angle\!\Big\langle
\operatorname{conj}(c_0)c_1
\Big\rangle_{\mathrm{broadside}},
\label{eq:phw}
\end{equation}
where the average is taken with the UE positioned at broadside, where the geometric phase difference is zero, so the measured phase corresponds directly to $\phi_{\mathrm{hw}}$. The resulting calibration is specific to the USRP B210 (serial \texttt{30F7DC2}) and its RX2-A/B receive paths.\\
\noindent\textit{2) Calibration procedure:} The hardware offset was first measured using a conducted setup in which both receive channels were driven from a common source through a power splitter (Mini-Circuits ZC2PD-K0244+), eliminating propagation and multipath effects. The conducted measurements yielded offsets of $+16.91^\circ$ at n40 and $+29.38^\circ$ at n78, confirming the frequency dependence of $\phi_{\mathrm{hw}}$.
This calibration was then refined over the air (OTA) with the antenna array installed and the UE placed at broadside. For the n40 experiments, the final calibrated value was $
\phi_{\mathrm{hw}} = +14.5^\circ,
$
subsequently applied to all AoA estimates; the difference from the conducted measurement reflects additional phase contributions from the antennas, mounting structure, and free-space propagation, absent in the cable measurement.
The calibrated value $\phi_{\mathrm{hw}}=+14.5^\circ$ was used without modification throughout both the indoor and outdoor campaigns, since the entire setting- the antenna array, cabling, and USRP hardware-  remained unchanged. Repeated measurements after retuning and power cycling confirmed that the broadside AoA consistently returned to approximately $90^\circ$, indicating negligible hardware phase drift; recalibration is only required if the RF hardware or cabling is modified.\\
\noindent\textit{3) Residual Systematic Error:}
Compensation of $\phi_{\mathrm{hw}}$ removes the constant inter-channel phase offset but not the residual angle-dependent AoA bias observed away from broadside. This bias is primarily caused by propagation effects, particularly multipath, and varies with the measurement environment: its sign changes between the indoor and outdoor campaigns, indicating that it is dominated by scene-dependent scattering rather than hardware calibration, with mutual coupling between the half-wavelength-spaced antenna elements a possible additional contributor. 
\subsection{MUSIC-based AoA estimation}
\label{music_music}
The proposed AoA estimator operates on the native per-antenna SRS channel estimates provided by the srsRAN gNB. For each SRS occasion, the complex channel coefficients from the two receive antennas are denoted by $c_0$ and $c_1$. Their relative phase,
$\Delta\varphi=\angle\langle c_0c_1^*\rangle$\label{diff_eqq}, is obtained by averaging over the SRS resource elements. For a two-element ULA with wavelength $\lambda$ and spacing $d=\lambda/2$, the phase difference is related to the AoA by~\cite{wachowiak2022angle}
\begin{equation}
\Delta\varphi=\frac{2\pi d}{\lambda}\cos\theta_{\mathrm{axis}},
\label{eq:geom}
\end{equation}
where $\theta_{\mathrm{axis}}$ is the AoA measured from the array axis.
The MUSIC algorithm forms the covariance matrix~\cite{wachowiak2022angle}
\begin{equation}
\mathbf{R}=\frac{1}{M}\mathbf{X}\mathbf{X}^{H},
\label{eq:cov}
\end{equation}
where $M$ is the number of received samples and $\mathbf{X}=[c_0;c_1]$ contains the SRS snapshots. The noise-subspace eigenvector $\mathbf{E}_n$ of $\mathbf{R}$ is then used to evaluate the MUSIC pseudospectrum~\cite{wachowiak2022angle}
\begin{equation}
P(\theta)=\frac{1}{\left|\mathbf{a}^{H}(\theta)\mathbf{E}_n\right|^{2}},
\label{eq:music}
\end{equation}
using the steering vector $\mathbf{a}(\theta)=\left[1,\ e^{j2\pi(d/\lambda)\sin\theta}\right]^T$. The AoA estimate corresponds to the angle that maximizes $P(\theta)$.
Since the proposed receiver employs only two antennas, MUSIC reduces to the classical phase-interferometer solution and provides neither super-resolution nor multipath separation. The closed-form estimate,
\begin{equation}
\theta_{\mathrm{coarse}}
=\arcsin\!\left(\frac{-\lambda\Delta\varphi}{2\pi d}\right),
\label{eq:coarse}
\end{equation}
is used to initialize the search, which is refined by maximizing the MUSIC pseudospectrum within a $\pm15^\circ$ window and the reported angle is $\theta_{\mathrm{axis}}=90^\circ-\theta$. MUSIC is applied independently to each UE-specific SRS channel estimate, yielding a single-source AoA estimate from the composite inter-element phase. Consequently, multipath arrivals are not separately resolved but instead manifest as a bias in the estimated AoA (see Algorithm~\ref{alg:music}).

\section{Measurement Methodology and Experimental Evaluation}\label{Measurment_real}

The proposed passive AoA system was evaluated through complementary indoor (see Fig.~\ref{fig:setup2}) and outdoor (see Fig.~\ref{fig:outdoor_setup}) measurement campaigns.
The corresponding measurement setups and procedures are described below.

\subsection{Indoor Measurements}\label{indoor_indoor}

Campaign~1 establishes the indoor baseline for evaluating the proposed AoA estimation method. 
The measurements were conducted in two rooms on the UPV campus using the n40 and n78 bands,
evaluating single-UE AoA accuracy over a broadside angular sweep, 
simultaneous AoA estimation of multiple handsets using code-orthogonal SRS, 
the effect of UE-gNB distance on AoA performance,
and the impact of operating frequency on link quality and throughput.

\noindent\textit{1) Indoor measurement setup and procedure:}
The indoor campaign was conducted in two environments (Fig.~\ref{fig:setup2}). 
Setup~1, a laboratory room, was used for the single-UE accuracy measurements, 
whereas Setup~2 was used for the multi-UE, distance-dependent, and n78 experiments. 
Table~\ref{tab:geometry} summarizes the measurement geometry. 
The array was mounted at 1.5\,m, 
with a nominal UE distance of 2\,m. 
UE1 and UE2 were positioned at array height, 
while UE3 was placed 0.5\,m lower, 
corresponding to a $14^\circ$ elevation angle. 
For a horizontal two-element array,
the measured phase scales as $\cos\theta_{\mathrm{az}}\cos\theta_{\mathrm{el}}$,
resulting in an approximately 3\% azimuth compression at $14^\circ$ elevation.

For each experiment,
the UE(s) were positioned at the desired AoA,
uplink \texttt{iperf3} traffic was initiated, 
and SRS occasions were extracted from the gNB log,
while the UE application logged the reference angle provided by the goniometer (Section~\ref{music_music}).
Individual UEs were distinguished by their RNTIs and orthogonal comb-4 SRS sequences,
allowing one AoA estimate per UE from the corresponding channel estimate. 
A single hardware phase calibration, $\phi_{\mathrm{hw}}=+14.5^\circ$ (Section~\ref{sec:calib}), was used throughout the campaign. 
In total, the dataset comprises approximately 1.55 million SRS occasions at n40 and 352\,000 at n78. 
The indoor environment also contained a commercial picocell operating in the n78 band, 
introducing strong co-channel interference, 
and Wi-Fi transmissions in the adjacent 2.4\,GHz band, close to n40.
The higher-power picocell significantly elevated the interference floor on n78,
whereas Wi-Fi had a comparatively smaller impact on n40.
Consequently, the received SRS SINR was substantially lower on n78, resulting in noisier AoA estimates than on n40.

\begin{figure}[t]
    \centering
    
    \begin{subfigure}{0.9\columnwidth}
        \centering
        \includegraphics[width=\linewidth]{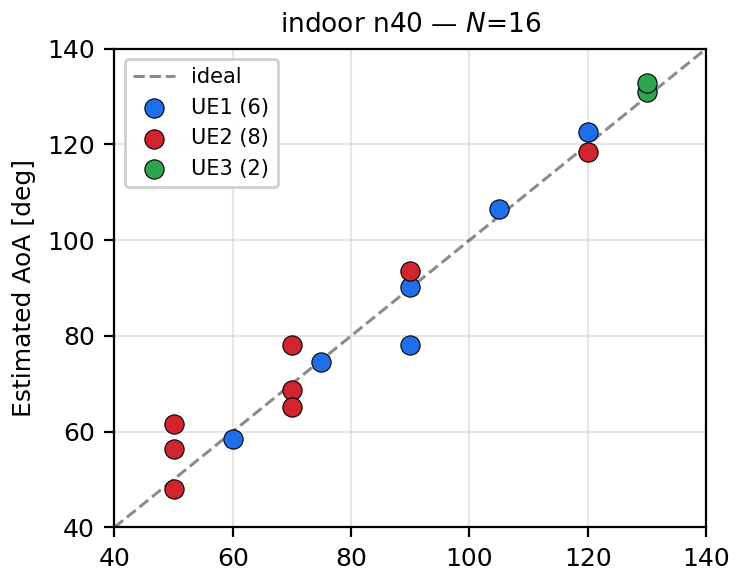}
        \caption{n40: all $N=16$ set-points.}
        \label{fig:estvtrue_n40}
    \end{subfigure}
    
    \vspace{0.6em}
    
    \begin{subfigure}{0.9\columnwidth}
        \centering
        \includegraphics[width=\linewidth]{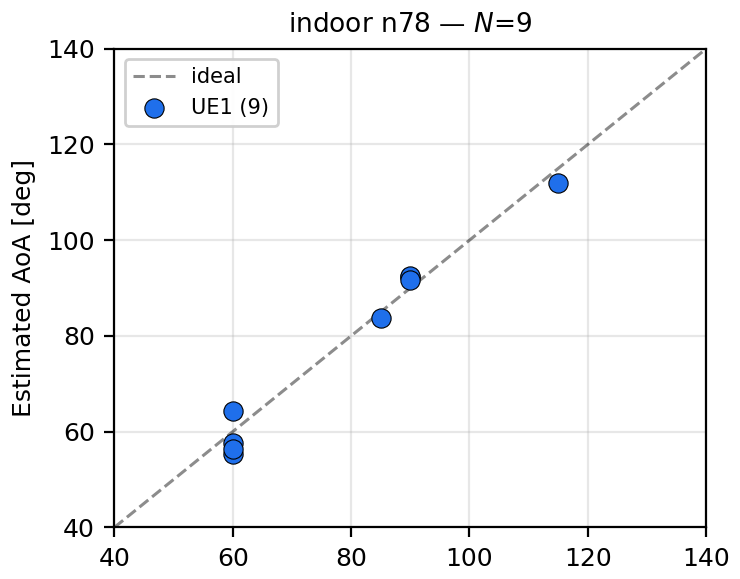}
        \caption{n78: all $N=9$ set-points.}
        \label{fig:estvtrue_n78}
    \end{subfigure}
    
    \caption{Estimated versus true AoA over every indoor measurement set-point, with one colour per handset.}
    \label{fig:estvtrue}
\end{figure}

\noindent\textit{2) AoA accuracy:}
For evaluating performance, 
we compare the estimated and true AoA for all indoor measurement set-points at n40 (Figure.~\ref{fig:estvtrue} a) and n78 (Figure.~\ref{fig:estvtrue} b). 
The dashed line represents the ideal $y=x$ relationship.

\begin{figure*}[t]
    \centering
    \includegraphics[width=0.9\textwidth]{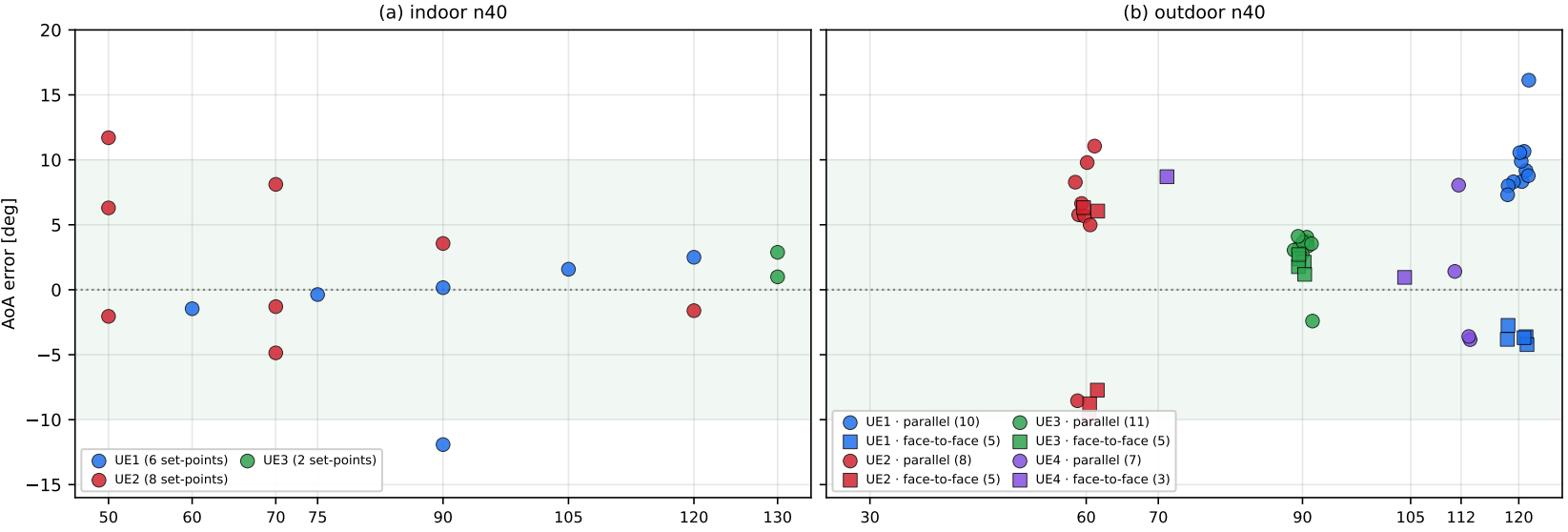}
    \caption{AoA error versus true AoA. (a) Indoor ($N=16$). (b) Outdoor ($N=54$) in the parallel and face-to-face scenarios.}
    \label{fig:exp2}
\end{figure*}

\begin{figure}[t]
    \centering
    \includegraphics[width=0.9\columnwidth]{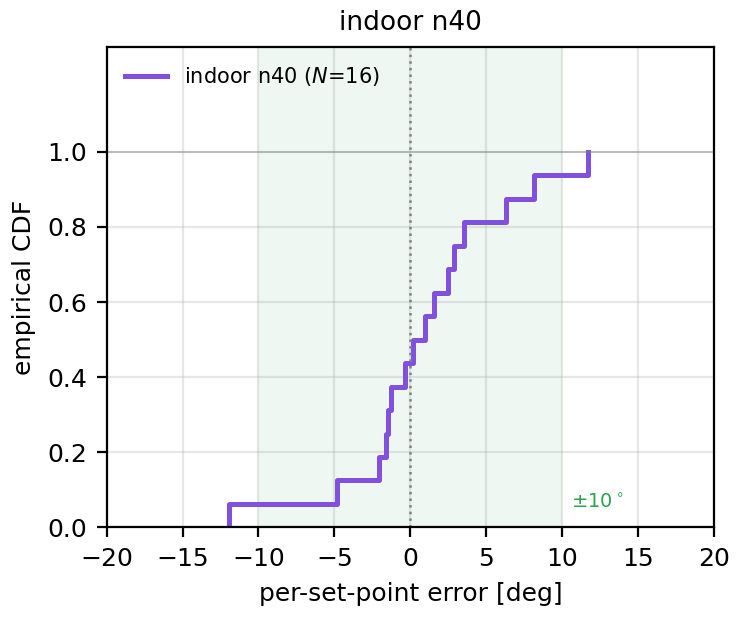}
    \caption{Empirical CDF of the per-set-point AoA error.}
    \label{fig:errcdf_indoor}
\end{figure}

For n40, the 16 indoor set-points closely follow the identity line over the \(60^\circ\)-\(120^\circ\) range, 
with a mean absolute error of $3.7^\circ$ and a maximum absolute error of $12^\circ$. 
The nine n78 set-points remain similarly aligned despite the lower SRS SINR, 
with corresponding errors of $2.8^\circ$ and $5^\circ$, respectively. 
Thus, although the per-occasion estimates are noisier at n78, 
averaging over the SRS occasions preserves an accurate set-point estimate, 
indicating that the calibration remains effective across the two operating bands.
The corresponding AoA error as a function of the true AoA for n40 is shown in Figure.~\ref{fig:exp2}. 
The error is smallest near broadside and increases toward endfire, 
consistent with the reduced angular sensitivity of the two-element baseline and the greater conversion of a given phase perturbation into angular error away from broadside.
Indoor multipath further contributes to the residual systematic bias because the two-element receiver cannot resolve individual propagation paths.
Table~\ref{tab:err} further collects the per-configuration error statistics for n40.
Although each set-point contains a large number of SRS occasions,
these samples are temporally correlated and therefore do not represent independent angular observations. 
Their low within-set-point variability and phasor concentration \(R>0.99\) nevertheless demonstrate high measurement precision. 
Consequently, the residual error is dominated mainly by systematic set-point bias rather than random estimation jitter. 
We therefore use the distinct angular set-points as the effective samples, 
report the mean error as the primary statistic,
and omit RMS values where only \(N=2\) independent set-points are available.
Figure~\ref{fig:errcdf_indoor} complements these statistics by showing the per-set-point errors and their empirical CDF.

\begin{table}[tp]
    \centering
    \caption{Per-configuration indoor AoA error with various set-points ($N$).}
    \label{tab:err}
    \footnotesize
        \begin{tabular}{@{}lllrrrr@{}}
        \toprule
        Config. & Band & UE(s) & $N$ & RMS & Bias & $|$Err$|_{\max}$ \\
        \midrule
        Single-UE sweep & n40 & UE1     & 5  & 1.5 & $+0.5$ & 2.5 \\
        Multi-UE (3$\to$1) & n40 & UE1-3 & 6  & 6.2 & $+0.5$ & 10.5 \\
        Alone$^{\dagger}$     & n40 & UE2     & 2  & - & $+1.3$ & 3.5 \\
        Distance$^{\dagger}$ & n40 & UE2     & 3  & 8.6 & $-2.0$ & 12.5 \\
        \bottomrule
        \end{tabular}  
\end{table}



\begin{figure*}[t]
    \centering
    \includegraphics[width=0.85\textwidth]{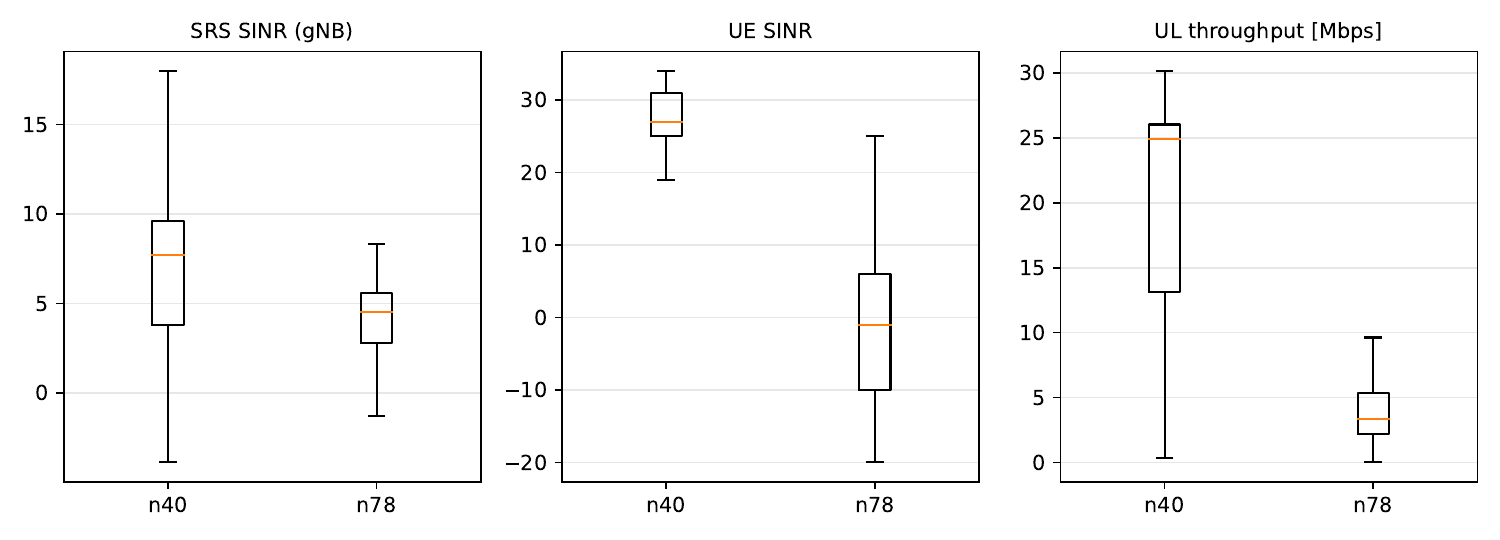}
    \caption{n40 vs.\ n78 distributions of SRS SINR, UE SINR and uplink throughput for UE1.}
    \label{fig:n40n78}
\end{figure*}
\begin{figure}[t]
    \centering
    \includegraphics[width=.9\columnwidth]{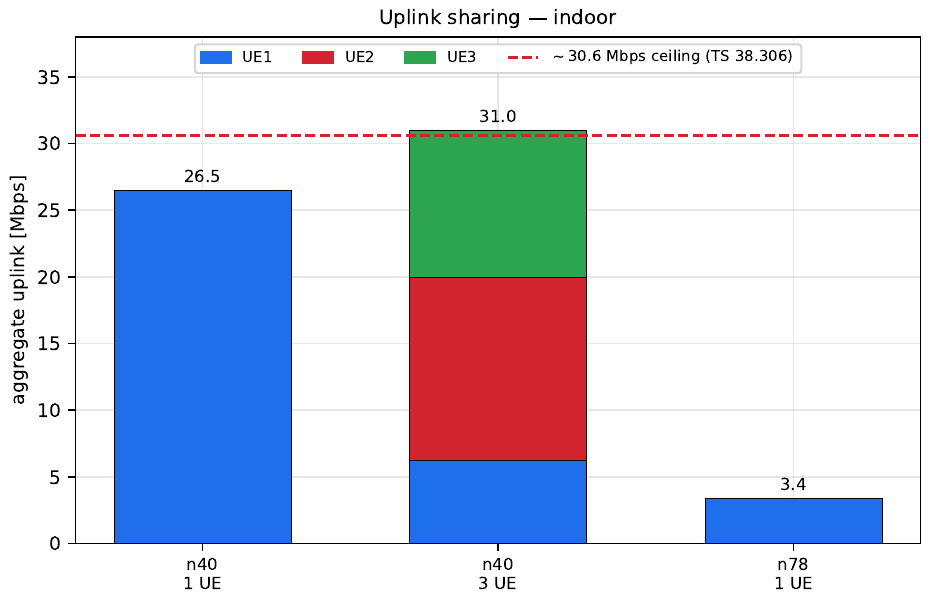}
    \caption{Measured uplink throughput for n40 and n78 compared with the practical TS~38.306 uplink limit.}
    \label{fig:thr}
\end{figure}

\noindent\textit{3) n40 vs.\ n78 and uplink performance:}
Figure~\ref{fig:n40n78} compares the two bands over the whole campaign,
while Table~\ref{tab:band} summarizes their median indoor link quality and uplink performance.
n78 performs markedly worse indoors,
with UE SINR about 28\,dB lower and uplink throughput about $7\times$ lower than n40.
Since the same antenna type is used in both bands, 
the degradation is instead associated primarily with the less favorable propagation conditions at 3.6 GHz and,
in particular, the strong co-channel interference from the indoor n78 picocell.
Figure~\ref{fig:thr} provides a complementary view of the uplink throughput relative to the practical TS~38.306 ceiling of approximately 30.6\,Mbps.
For n40, a single UE achieves about 26.5\,Mbps ($87\%$ of this limit),
while three simultaneous UEs reach an aggregate throughput of 31.0\,Mbps,
indicating efficient utilization of the available uplink resources.
The unequal throughput allocation among simultaneous UEs is primarily due to scheduler contention.
In contrast, the n78 measurements, limited to a single UE, achieve only 3.4\,Mbps, 
confirming the strong throughput penalty associated with the much weaker indoor link at 3.6\,GHz.
AoA quality generally follows link quality across the campaign,
but the two are not interchangeable.
In particular, the distance-dependent measurements show that a robust data link can coexist with degraded AoA:
throughput requires sufficient post-combining SINR,
whereas AoA estimation additionally depends on preserving the spatial phase structure across the array,
which is more sensitive to multipath.
Consequently, communication link quality or throughput alone cannot be used as a proxy for AoA accuracy.


\begin{table}[tp]
    \centering
    \caption{Indoor band summary (per-occasion medians)}
    \label{tab:band}
    \footnotesize
        \begin{tabular}{@{}lrrrr@{}}
        \toprule
        Band & SRS SINR (dB) & UE SINR (dB) & UL (Mbps) & RSRP (dBm) \\
        \midrule
        n40 & 7.7 & 27.0 & 24.9 (30.2) & $-94$ \\
        n78 & 4.5 & $-1.0$ & 3.4 (10.3) & $-99$ \\
        \bottomrule
        \end{tabular}
\end{table}

\subsection{Outdoor Measurement}\label{outdoor_outdoor}

\begin{figure*}[!htb]
    \centering
    \includegraphics[width=0.87\textwidth]{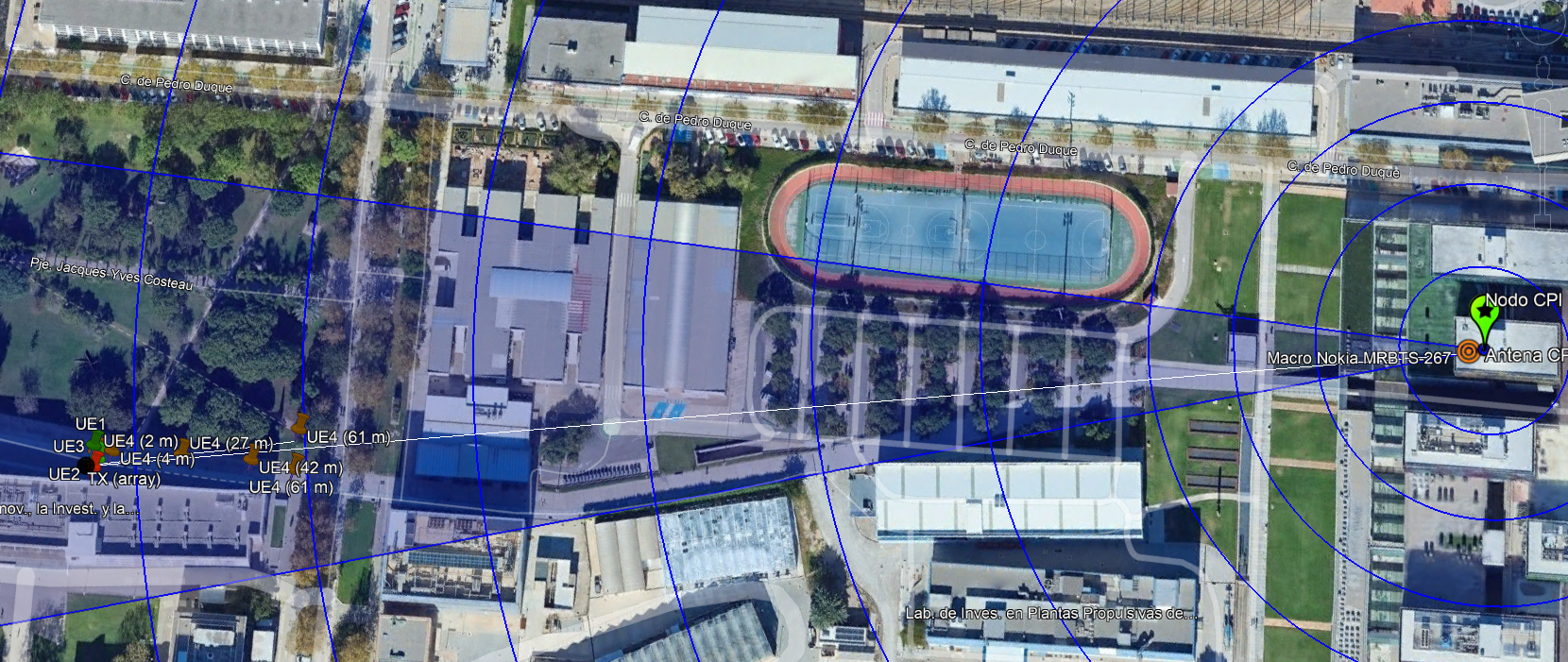}
    \caption{Georeferenced outdoor propagation geometry. The $\sim110\degr$ radial runs into the park
    (foliage-shadowed, clean); the $\sim80$-$85\degr$ radial runs down the inter-building corridor
    (strong scattering). The Nokia macro \texttt{MRBTS-267} (409\,m, AoA $102\degr$) marks the
    corridor's direction but is an adjacent-channel emitter, not the cause of the AoA failure.}
    \label{fig:mapa}
\end{figure*}

The second measurement campaign extends the indoor evaluation to a real, uncontrolled outdoor propagation environment on the UPV campus, in front of the Nexus building (see Fig.~\ref{fig:mapa}). The area is characterized by extensive glass façades and surrounding structures that create strong reflections and a rich multipath environment. The campaign was performed on n40 only, using a 20\,MHz bandwidth and 30\,kHz subcarrier spacing.
We estimated the AoA of up to four COTS 5G handsets simultaneously, with each UE processed independently from its own uplink SRS channel estimate using comb-4 and RNTI-based separation, as in the indoor campaign. The same two-element USRP B210 array and the same hardware phase calibration, $\phi_{\mathrm{hw}}=+14.5^\circ$, were used throughout the campaign. The system provides one AoA estimate per UE over an approximately $60^\circ$ usable angular sector rather than a full 2D position estimate. Uplink \texttt{iperf3} traffic was active throughout the measurements, allowing AoA, SRS SINR, and uplink throughput to be recorded jointly for each UE.
\noindent\textit{1) Outdoor measurement setup and scenarios:}
Two measurement geometries were considered.
In the \emph{parallel} geometry, the array and UEs were positioned side by side and oriented in the same direction, with the UEs placed at a height of 1.5\,m (Fig.~\ref{fig:outdoor_setup_a}).
In the \emph{face-to-face} geometry, the array faced the Nexus glass façade, with the UEs positioned at a height of 2\,m (Fig.~\ref{fig:outdoor_setup_b}).
This second configuration exposed the received uplink signals to stronger reflections and scattering from the glass façade and surrounding structures.
Four Samsung handsets were used: UE1 (S23), UE2 (S25 Ultra), UE3 (S25 Ultra), and UE4 (S23).
The experiments covered simultaneous AoA estimation of up to four UEs, fixed-angle measurements, distance-dependent measurements, and mobility tracking.
For the fixed-angle measurements, UE1, UE2, and UE3 were positioned at $120^\circ$, $60^\circ$, and $90^\circ$, respectively, while UE4 served as the moving handset.
For the distance-dependent measurements, UE4 was positioned at 2, 4, 27, 42, and 61\,m from the array.
A separate mobility experiment tracked UE4 while it moved across the full $0^\circ$--$180^\circ$ angular range.

\begin{figure}[t]
    \centering
    \includegraphics[width=0.8\columnwidth]{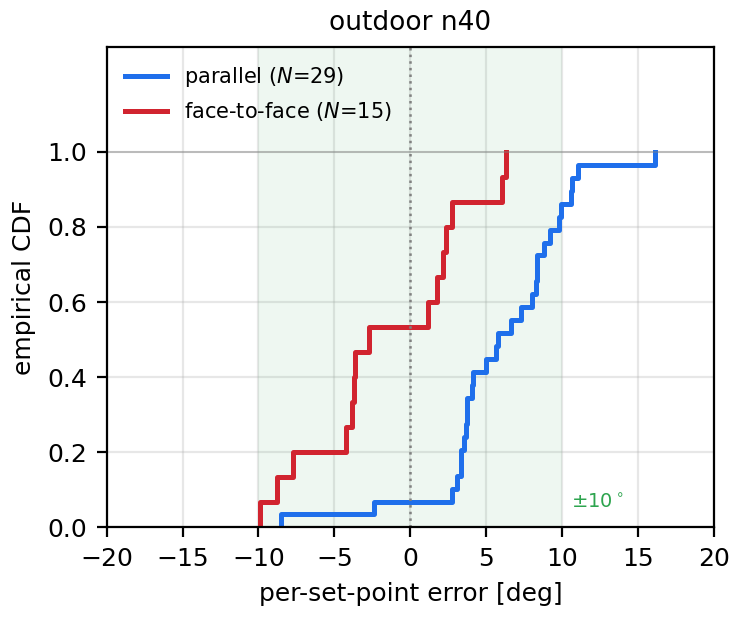}
    \caption{Empirical CDF of the per-set-point AoA error.}
    \label{fig:errcdf_outdoor}
\end{figure}

\begin{figure}[t]
    \centering
    \includegraphics[width=0.39\textwidth]{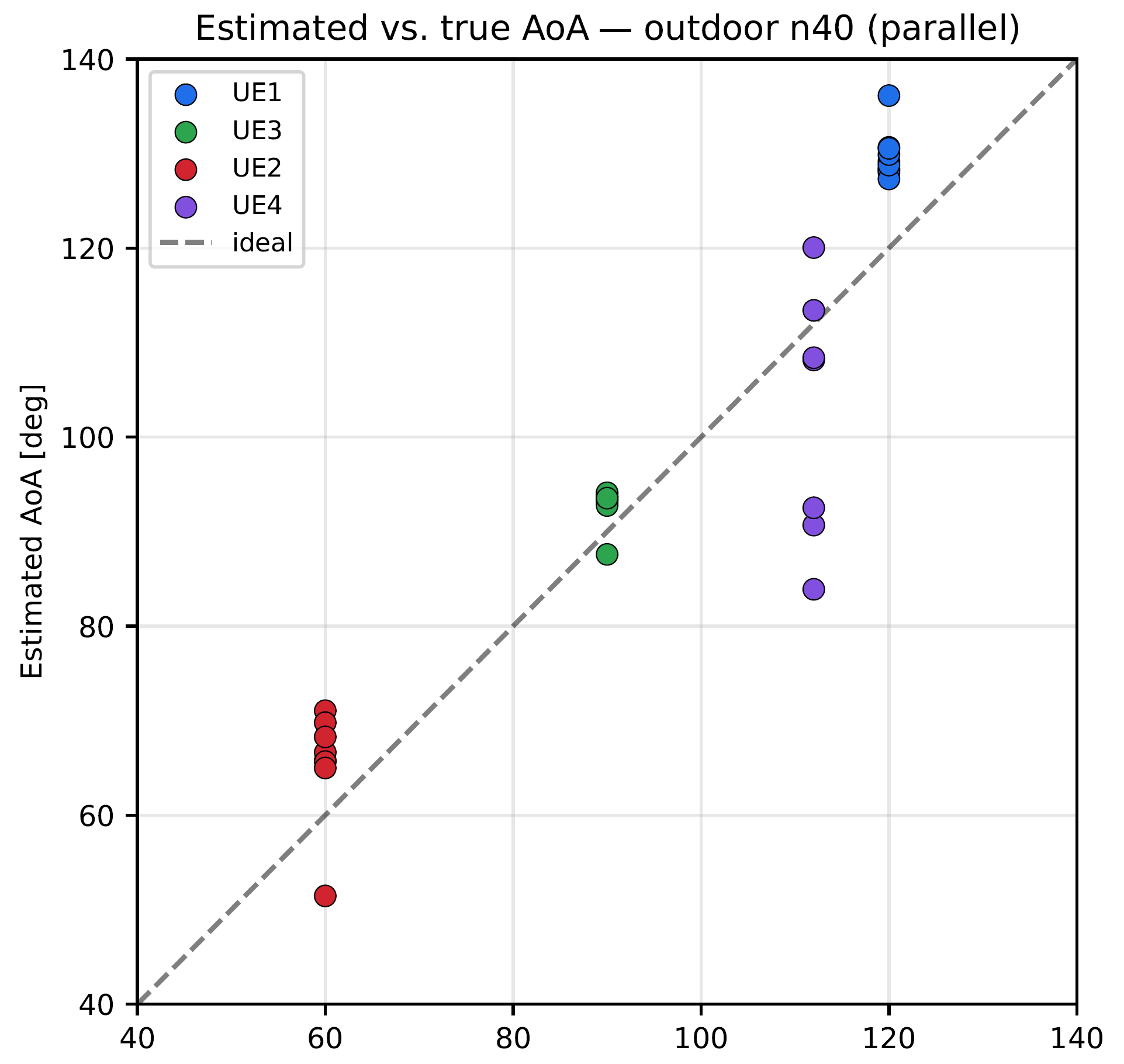}
    \caption{Estimated vs.\ true AoA outdoors (Four UEs).}
    \label{fig:estvout}
\end{figure}

\noindent\textit{2) AoA accuracy and geometry-dependent bias:}
Outdoor AoA estimates roughly track the true angle but show a clear positive bias and larger scatter than indoors.
Figure~\ref{fig:estvout} compares the estimated and true AoA for the outdoor measurement set-points, while Table~\ref{tab:outacc} summarizes the corresponding per-configuration error statistics. The estimates generally follow the ideal $y=x$ relationship, but exhibit systematic biases that depend strongly on the measurement geometry. For the parallel configuration (n40), the mean absolute error is approximately $9$--$10^\circ$, with a maximum of $27$--$28^\circ$ driven by an outlier from UE4, reflecting the increased multipath and non-line-of-sight propagation typical of outdoor environments. Figure~\ref{fig:errcdf_outdoor} complements these statistics by showing the empirical distribution of the per-set-point AoA errors.
In the \emph{parallel} geometry, UE1, UE2, and UE3 exhibit mean AoA errors of $+9.7^\circ$, $+5.5^\circ$, and $+3.0^\circ$, respectively. In the \emph{face-to-face} geometry, the corresponding mean errors become $-3.6^\circ$, $-2.8^\circ$, and $+2.0^\circ$. UE1 and UE2 thus reverse the sign of their bias between the two geometries despite using the same array and hardware phase calibration, indicating that the dominant residual error is associated with scene-dependent multipath rather than a fixed calibration offset.
The outdoor reference AoAs are obtained from the goniometer (Section~\ref{music_music}), with a placement and read-off uncertainty of approximately $\pm1$-$2^\circ$. The observed biases therefore exceed this reference uncertainty in several configurations. The per-set-point phasors remain highly concentrated ($R>0.99$), indicating high within-set-point precision despite the residual absolute error. As in the indoor campaign, the many SRS occasions within a set-point are temporally correlated and do not constitute independent angular observations; we therefore treat distinct set-points as the effective samples for the error statistics.
The $95\%$ confidence intervals in Table~\ref{tab:outacc} further quantify variability across set-points. Those for UE1 and UE3 remain bounded away from zero in both geometries, whereas the UE2 face-to-face configuration shows substantially larger variability ($\sigma=8.3^\circ$, $N=5$) with a confidence interval of $[-13.1^\circ,+7.4^\circ]$, not statistically distinguishable from zero bias.
Overall, the mean bias stays within approximately $\pm10^\circ$ across the evaluated outdoor configurations, though individual set-points occasionally exceed this range. The change in both magnitude and sign of the error between geometries indicates that the dominant limitation of the two-element estimator outdoors is scene-dependent propagation rather than measurement instability.

\begin{table}[tp]
    \centering
    \caption{Outdoor per-configuration AoA error (n40).}
    \label{tab:outacc}
    \footnotesize
    \begin{tabular}{@{}lrrcrr@{}}
    \toprule
    UE & GT & Mean err. & $95\%$ CI & $\sigma$ & $N$ \\
    \midrule
    \multicolumn{6}{@{}l}{\emph{Parallel geometry (open park)}}\\
    UE1 & 120 & $+9.7$ & $[+7.9,\,+11.5]$  & 2.5 & 10 \\
    UE2 & 60  & $+5.5$ & $[+0.4,\,+10.5]$  & 6.0 & 8  \\
    UE3 & 90  & $+3.0$ & $[+1.8,\,+4.2]$   & 1.8 & 11 \\
    \addlinespace
    \multicolumn{6}{@{}l}{\emph{Face-to-face (Nexus glass facade)}}\\
    UE1 & 120 & $-3.6$ & $[-4.3,\,-3.0]$   & 0.5 & 5  \\
    UE2 & 60  & $-2.8$ & $[-13.1,\,+7.4]$  & 8.3 & 5  \\
    UE3 & 90  & $+2.0$ & $[+1.3,\,+2.8]$   & 0.6 & 5  \\
    \bottomrule
    \end{tabular}
    \end{table}
    
\begin{figure*}[t]
\centering
\includegraphics[width=0.9\textwidth]{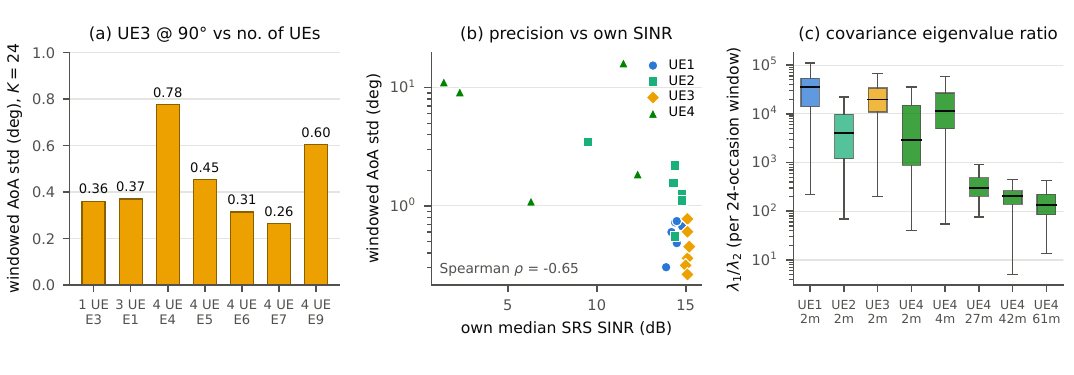}
\caption{Outdoor multi-UE statistics (Campaign~2, n40).}
\label{fig:multiuestats}
\end{figure*}

\noindent\textit{3) Multi-UE AoA and scalability:}
We next evaluate whether the simultaneous transmission of multiple UEs degrades AoA estimation performance. Three and subsequently four handsets were operated simultaneously, with the gNB providing an independent SRS channel estimate for each UE through comb-4 subcarrier interleaving and RNTI-based separation. The same two-element AoA estimator was then applied independently to each per-UE channel estimate.
Figure~\ref{fig:multiuestats} summarizes the AoA precision across experiments E1--E9. Each run was divided into non-overlapping windows of $K=24$ SRS occasions, yielding one AoA estimate per window, and the standard deviation of the windowed estimates was used as the per-UE precision metric.
Figure~\ref{fig:multiuestats}(a) shows that increasing the number of simultaneous UEs does not materially degrade AoA precision. For UE3, fixed at $90^\circ$, the windowed standard deviation is $0.36^\circ$ when operating alone and $0.37^\circ$ with three simultaneous UEs. Across the five four-UE experiments, it remains between $0.26^\circ$ and $0.78^\circ$, with a mean of $0.48^\circ$. Its median SRS SINR remains close to 15\,dB, and similar behavior is observed for the other fixed UEs. These results indicate that the orthogonal SRS resources effectively isolate the per-UE channel estimates.
In contrast, Fig.~\ref{fig:multiuestats}(b) shows that AoA precision is strongly related to each UE's own SRS quality. As the moving UE4 progresses from 2 to 61\,m, its median SRS SINR decreases overall from 11.5 to 2.3\,dB, while the standard deviation of its AoA estimates increases from approximately $1^\circ$ to $9^\circ$--$11^\circ$. The nearby fixed UEs maintain comparatively stable SINR and AoA precision. Across all runs, the windowed AoA standard deviation is negatively correlated with the median SRS SINR, with Spearman $\rho=-0.65$ and Pearson $r=-0.73$.
Figure~\ref{fig:multiuestats}(c) further shows the evolution of the covariance eigenvalue ratio $\lambda_1/\lambda_2$, used as a per-UE signal-subspace quality indicator. Although the ratio decreases by approximately 22\,dB as UE4 moves from 2 to 61\,m, it remains above unity, indicating that a usable signal subspace is preserved even at the largest evaluated distance.
These results show that the observed loss of AoA precision is primarily associated with the quality of each UE's own SRS channel estimate rather than with the number of simultaneously active UEs. Importantly, the two-element array does not spatially separate the UEs: user separation is provided by the orthogonal SRS resources before the AoA estimator is applied independently to each UE. Figure~\ref{fig:guiout} provides representative examples of the live three- and four-UE operation.
The four-UE experiment represents the implemented configuration rather than a fundamental scalability limit. Additional NR SRS multiplexing resources, including frequency-domain offsets, cyclic shifts, and time-division multiplexing, can support larger UE counts subject to the available SRS resources and gNB configuration.

\begin{figure}[t]
\centering
\includegraphics[width=1\columnwidth]{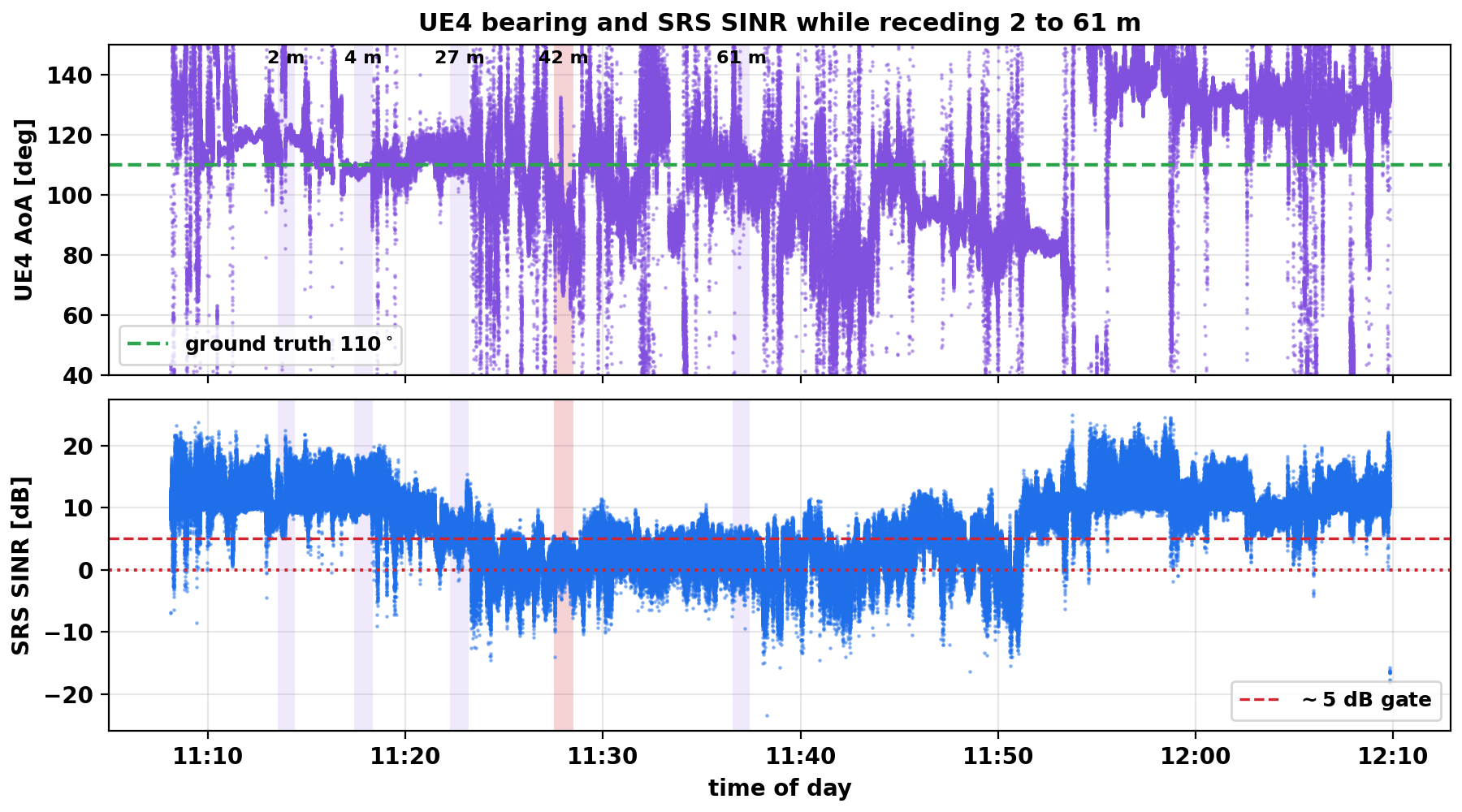}
\caption{UE4 AoA (top) and SRS SINR (bottom) over time.}
\label{fig:ue4ts}
\end{figure}

\noindent\textit{4) Distance-dependent performance:}
We next investigate how AoA performance evolves with UE--array distance and SRS link quality. Figure~\ref{fig:exp4} compares the indoor and outdoor distance-dependent measurements. Indoors, UE2 was kept at approximately $70^\circ$ while its distance from the array was increased from 2 to 6\,m. Outdoors, UE4 was evaluated at 2, 4, 27, 42, and 61\,m.
The indoor measurements show a gradual degradation in AoA performance across the evaluated distances, with an RMS AoA error of $8.6^\circ$ across the three distance set-points. Outdoors, however, the behavior is clearly non-monotonic: the largest AoA error occurs at 42\,m, while the estimate partially recovers at 61\,m despite the greater distance. This behavior shows that distance alone is not a reliable predictor of AoA accuracy in a realistic multipath environment.
The outdoor time series in Fig.~\ref{fig:ue4ts} further illustrates this behavior. UE4 exhibits stable AoA estimates at the shorter, higher-SINR positions, substantially larger fluctuations around 42\,m, and partial recovery at 61\,m. The corresponding SRS SINR also varies non-monotonically along the trajectory, indicating that the local propagation conditions, rather than distance alone, govern the quality of the received SRS and the resulting AoA estimate.
In contrast, uplink throughput decreases more consistently with distance outdoors, while remaining approximately 24--26\,Mbps over the evaluated indoor distances. This different behavior reinforces that communication performance and AoA accuracy are related through link quality but are not interchangeable metrics.

\begin{figure}[!b]
\centering
\includegraphics[width=0.85\columnwidth]{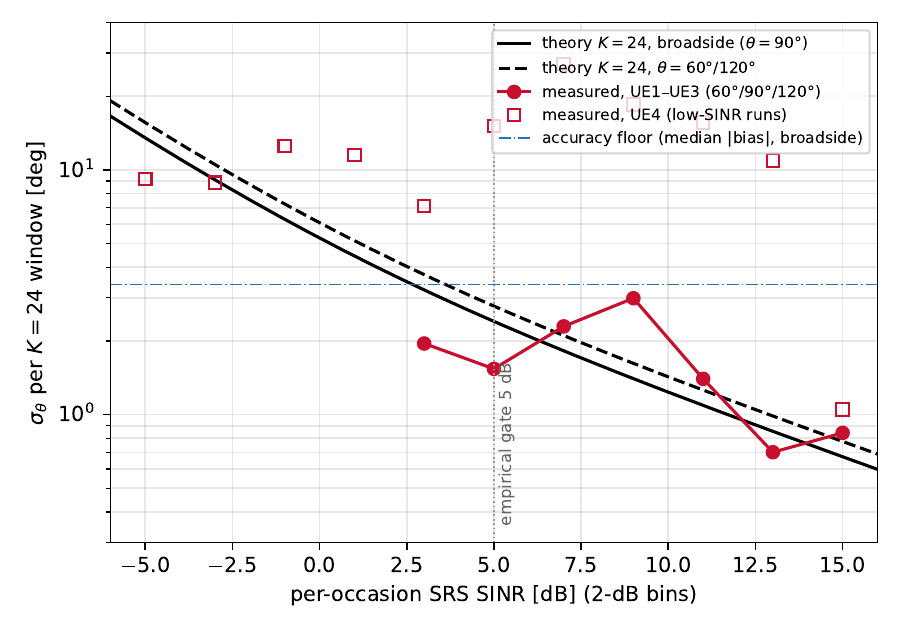}
\caption{AoA precision against the noise-only baseline of
Eqs.~\eqref{eq:vardphi}-\eqref{eq:sigtheta} for windows of $K=24$ SRS
occasions.}
\label{fig:crlb}
\end{figure}

\begin{figure*}[t]
\centering
\begin{subfigure}[b]{0.49\textwidth}
  \centering
  \includegraphics[width=\textwidth]{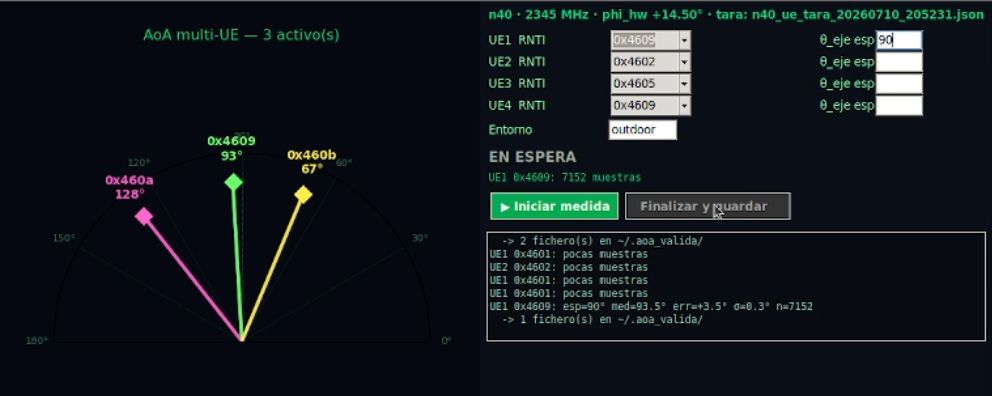}
  \caption{Three simultaneous UEs.}
\end{subfigure}
\hfill
\begin{subfigure}[b]{0.49\textwidth}
  \centering
  \includegraphics[width=\textwidth]{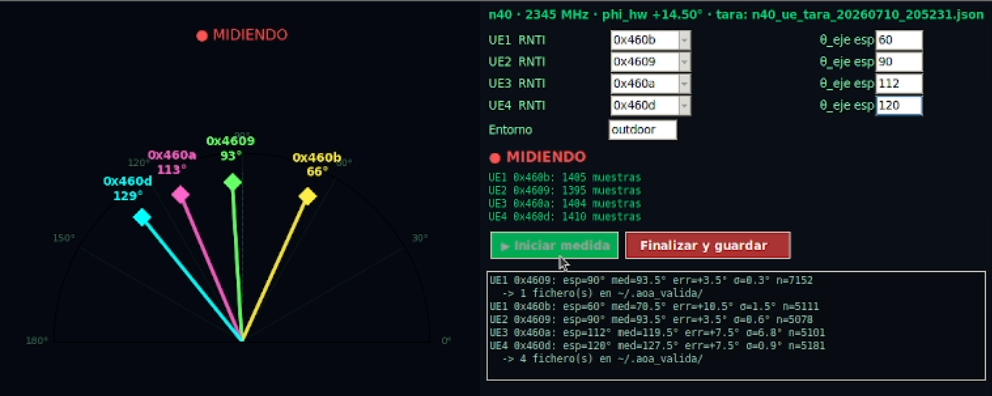}
  \caption{Four simultaneous UEs.}
\end{subfigure}
\caption{Live outdoor GUI showing three (left) and four (right) simultaneous per-UE AoA estimates. Each AoA is computed independently from the corresponding handset's RNTI- and comb-4-orthogonal SRS, rather than by spatial separation.}
\label{fig:guiout}
\end{figure*}
\begin{figure*}[t]
\centering
\includegraphics[width=0.92\textwidth]{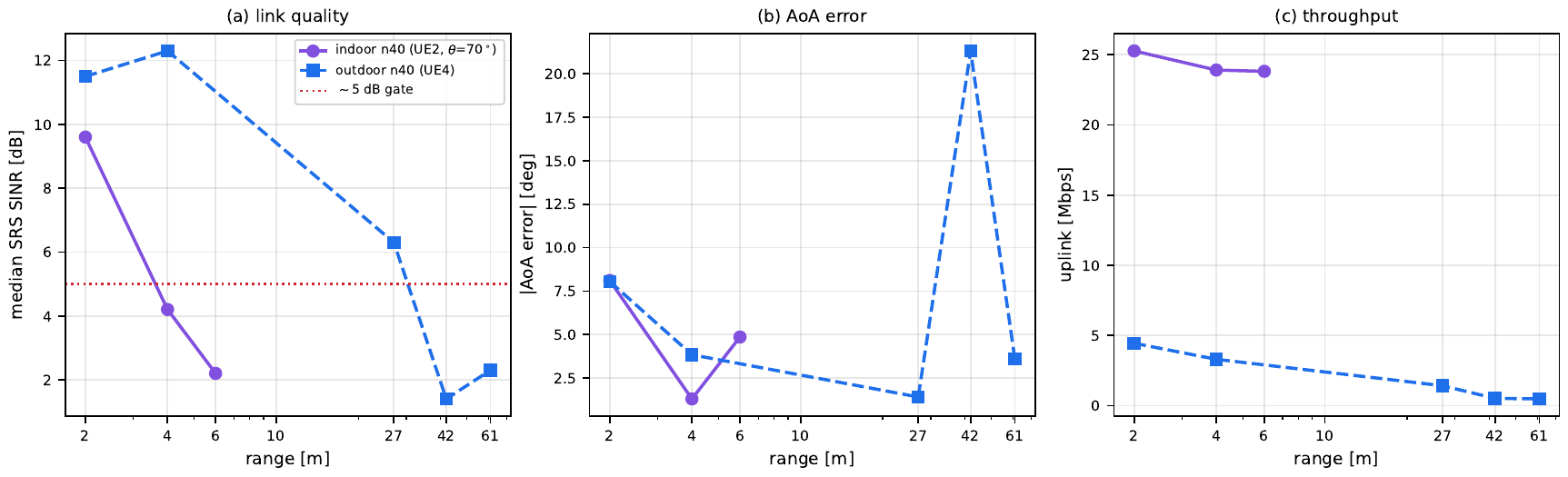}
\caption{Range sweep across both campaigns on a common logarithmic axis: indoor (UE2 at
$\theta=70\degr$, 2/4/6\,m) and outdoor (UE4, 2/4/27/42/61\,m).}
\label{fig:exp4}
\end{figure*}

\begin{figure}[t]
\centering
\includegraphics[width=0.92\columnwidth]{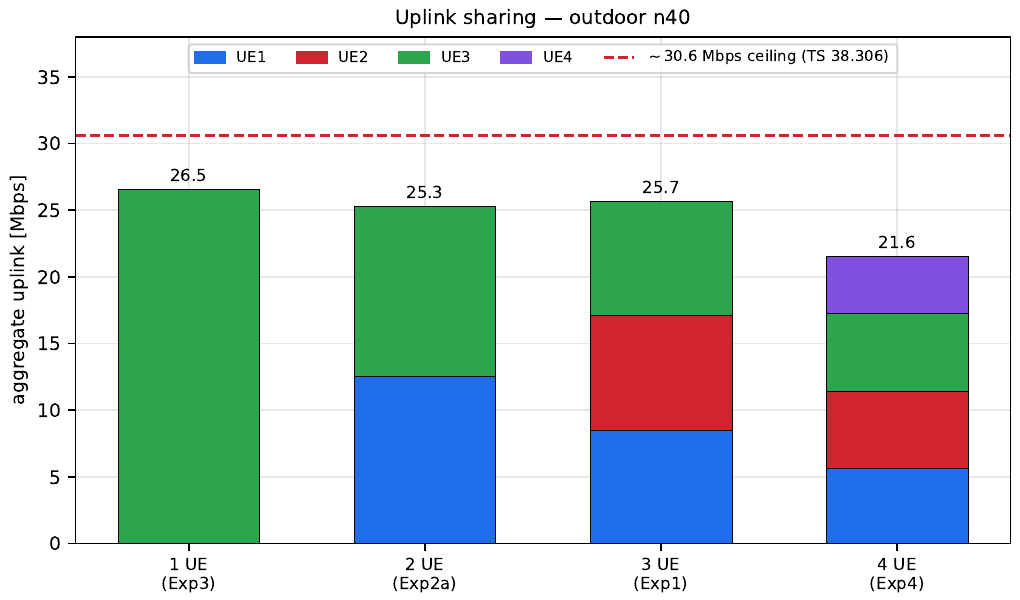}
\caption{Uplink throughput for 1-4 simultaneous UEs.}
\label{fig:sharingout}
\end{figure}

\begin{figure}[t]
\centering

    \begin{subfigure}{0.8\columnwidth}
        \centering
        \includegraphics[width=\linewidth]{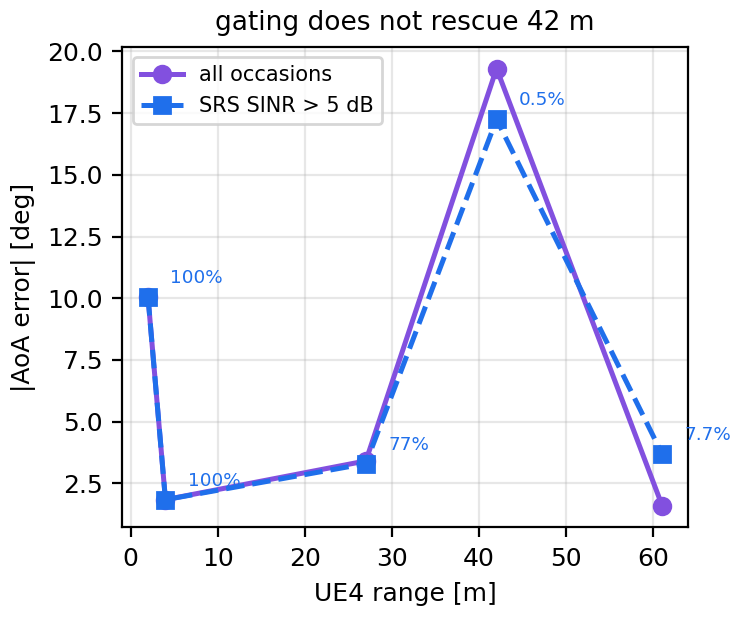}
        \caption{Walking trajectory of UE4.}
        \label{fig:ue4traj}
    \end{subfigure}
    
    \vspace{0.5em}
    
    \begin{subfigure}{0.8\columnwidth}
        \centering
        \includegraphics[width=\linewidth]{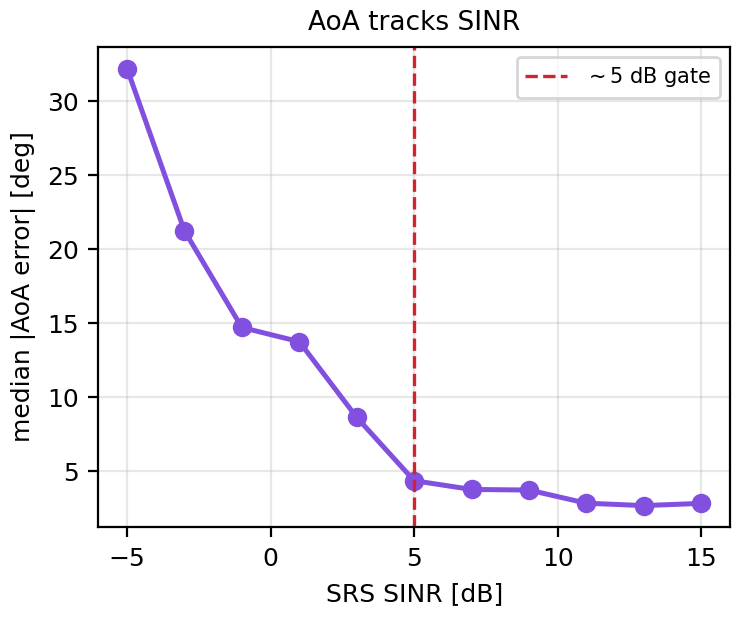}
        \caption{Absolute AoA error as a function of the SRS SINR.}
        \label{fig:ue4sinr}
    \end{subfigure}

\caption{Performance of the moving UE4.}
\label{fig:sinrgate}
\end{figure}

\noindent\textit{5) Mobility and usable angular sector:}
We evaluate AoA tracking under UE mobility. UE4 was moved continuously across the $0^\circ$--$180^\circ$ angular range while maintaining an active uplink transmission. The estimated AoA follows the UE trajectory most reliably within approximately $60^\circ$--$120^\circ$, whereas the estimates become increasingly unstable outside this region.
This behavior is consistent with the geometry of the two-element array. Angular sensitivity decreases toward endfire, where a given inter-antenna phase perturbation translates into a larger angular error. In addition, the two-element ULA exhibits a front--back ambiguity and therefore cannot uniquely distinguish arrivals from opposite sides of the array. The outdoor mobility experiment consequently confirms an effective operating sector of approximately $60^\circ$ around broadside.
Within this sector, the system can continuously track a moving COTS handset from its native uplink SRS without requiring any modification to the UE. Outside the sector, the reported AoA should be interpreted with caution because the array geometry, rather than the SRS processing itself, becomes the dominant limitation.\\
\noindent\textit{6) AoA sensitivity to noise, geometry, and propagation conditions:}
The preceding measurements show that AoA accuracy is jointly determined by SRS link quality, array geometry, and the propagation environment. 
The noise-limited baseline (Figure~\ref{fig:crlb}) predicts an AoA
error of $2.4^\circ$--$2.8^\circ$ at 5\,dB SRS SINR, closely matching
the measured $3.4^\circ$ floor in Figure~4 ($K=24$). Above 12\,dB,
measured RMS precision ($0.78^\circ$) is close to the predicted
$0.6$--$1.0^\circ$, with only modest excess variance from
calibration drift and multipath. The dominant high-SINR error is
instead systematic bias ($3.4^\circ$ at broadside, $8.3^\circ$ at
$60^\circ/120^\circ$), which longer averaging cannot reduce.
The angular dependence is consistent with the two-element array geometry. Sensitivity is highest near broadside and decreases toward endfire, where a given inter-antenna phase perturbation produces a larger angular error. Moreover, the two-element receiver cannot resolve individual propagation paths, so reflected and scattered components can bias the effective inter-antenna phase even at adequate SRS SINR.
The experimental results in Fig.~\ref{fig:sinrgate} are consistent with this behavior. Above approximately 5\,dB SRS SINR, the AoA error remains close to the noise-limited baseline, whereas substantially larger errors occur at lower SINR. However, the relationship is not one-to-one: the non-monotonic distance-dependent behavior reported above shows that similar link-quality conditions can still produce different AoA errors depending on the local propagation environment. Thus, SRS SINR is a useful indicator of AoA reliability, but an SINR threshold alone cannot identify all large-error conditions when scene-dependent multipath dominates the measured inter-antenna phase.
Finally, Table~\ref{tab:budget} shows that bounded setup uncertainties remain comparatively small, whereas configuration-dependent AoA biases reach approximately $3^\circ$--$10^\circ$. Together with the high repeatability at fixed positions and the geometry-dependent biases observed outdoors, this indicates that the dominant residual error under favorable link conditions is scene-dependent propagation rather than random estimation noise or measurement instability.
\begin{table}[tp]
    \centering
    \caption{Consolidated uncertainty budget.}
    \label{tab:budget}
    \footnotesize
    \setlength{\tabcolsep}{4pt}
    \begin{tabular}{@{}llll@{}}
    \toprule
    Source & Type & Magnitude & Comment \\
    \midrule
    Goniometer reference & B & $\pm1$-$2\degr$ & placement, read-off \\
    Set-point grid ($5\degr$) & B & $\pm2.5\degr$ & bound, not a std \\
    Elevation (UE3) & B & $\approx-3\%$ & $\cos14\degr$ compression \\
    Calibration residual & B & $<1\degr$ & OTA broadside tare \\
    Mutual coupling & B & unquantified & $d=\lambda/2$; off-broadside \\
    Scene multipath & A & $\sigma$ $0.5$-$8.3\degr$ & dominant; per config. \\
    Windowed precision & A & $\sigma$ $0.5$-$2.5\degr$ & in-scene; $R>0.99$ \\
    \bottomrule
    \end{tabular}
\end{table}

\subsection{Implementation limitations and lessons learned}
\label{sec:impllimitations}

The main implementation limitation is the use of a two-element array.
With only one spatial baseline, the estimator cannot distinguish the
direct path from reflected or scattered components and therefore
provides an effective AoA rather than a resolved direction of arrival
for individual paths. Mutual coupling and the reduced angular
sensitivity toward endfire can introduce additional angle-dependent
bias. These limitations are inherent to the compact two-element
architecture and cannot be eliminated by phase calibration alone.

A second practical limitation is the dependence on the propagation
environment. Passive SRS-based AoA should therefore not be regarded as
a purely range-dependent measurement. Its reliability depends on the
instantaneous channel quality, the geometry of the surrounding
environment, and the presence of reflected or scattered paths.
Hardware calibration provides high repeatability within a fixed
configuration, but it cannot compensate for scene-dependent multipath
bias. Consequently, calibration alone does not guarantee that the same
absolute AoA accuracy will be maintained when the deployment geometry
changes.

The main practical lesson is therefore that the proposed two-element
architecture provides a low-complexity and repeatable network-side AoA
solution, but its absolute accuracy in realistic environments is
ultimately limited by the available spatial information and the local
propagation conditions. More demanding localization scenarios would
benefit from additional antenna elements or spatial diversity to
improve multipath discrimination and angular robustness.

\section{Discussion and Conclusion}

We presented a passive AoA estimation framework for commercial 5G handsets that operates directly on the native per-antenna SRS channel estimates produced by an srsRAN gNB. 
The proposed system requires no protocol modifications or UE cooperation and enables simultaneous AoA estimation of up to four COTS devices using a low-cost two-element USRP~B210 receiver.

Experimental results demonstrate that accurate passive AoA estimation from native 5G NR signaling is feasible.
Indoor measurements achieved an RMS AoA error of $1.5^\circ$ over the $60^\circ$--$120^\circ$ sector in n40, 
while outdoor experiments demonstrated simultaneous AoA estimation and tracking of multiple commercial handsets under realistic propagation conditions. In the \emph{parallel} geometry, UE1, UE2, and UE3 exhibit mean AoA errors of $+9.7^\circ$, $+5.5^\circ$, and $+3.0^\circ$, respectively. In the \emph{face-to-face} geometry, the corresponding mean errors become $-3.6^\circ$, $-2.8^\circ$, and $+2.0^\circ$.
The results show that AoA accuracy depends on SRS link quality, array geometry, and scene-dependent multipath rather than on transmitter distance alone.
Errors generally increase as SRS SINR approaches and falls below approximately $5$\,dB,
although this should be regarded as an empirical operating region rather than a universal threshold.

The current implementation is limited by the two-element array, 
for which MUSIC effectively reduces to a phase interferometer without multipath resolution and with an approximately $60^\circ$ usable sector around broadside. 
The reported n78 results are qualitative because they rely on calibration performed at n40. 
Future work will investigate larger arrays, band-specific calibration, and more robust multipath-aware processing for challenging propagation environments.

\bibliographystyle{ieeetr}
\bibliography{References}

\end{document}